\documentclass[12pt]{article}
\input epsf
\usepackage{amssymb}
\usepackage{amsmath}
\makeatletter
\@addtoreset{equation}{section}
\makeatother

\def\be{\begin{equation}}
\def\ee{\end{equation}}

\def\ba{\begin{array}}
\def\ea{\end{array}}

\newcommand{\bea}{\begin{eqnarray}}
\newcommand{\eea}{\end{eqnarray}}
\def\H{{\bf H}}
\def\G{{\bf \Gamma}}

\begin{document}
\begin{titlepage}

\begin{flushright}
CERN-PH-TH/2006-116 \\
SU-ITP-2006-19     \\
LBNL-60487 \\
UCLA/06/TEP/19\\
hep-th/xxxxxxx
\end{flushright}
\vspace{.5cm}
%\vskip 1cm

\vfill

\begin{center}
{ \Large {\bf    Magic Supergravities,  $N= 8$ and  \\
\vskip 0.8 cm Black Hole Composites   }}

\vspace{20pt}

{\bf  Sergio Ferrara$^{1}$, Eric G. Gimon$^{2}$ and Renata Kallosh$^{3}$
 } \\

\vspace{20pt}

{\small

$^1$ Physics Department, Theory Unit, CERN, 1211 Geneva 23, Switzerland \\
\vspace{6pt}
INFN, Laboratori Nazionali di Frascati, Via Enrico Fermi, 40,00044 Frascati, Italy \\
\vspace{6pt}
 Department of  Physics \& Astronomy, University of California, Los
Angeles, CA 90095, USA \\ \vspace{6pt}
{\it Sergio.Ferrara@cern.ch}\\

\vspace{10pt}

$^2$  Department of Physics, University of California, Berkeley, CA 94720,
USA\\ \vspace{6pt} Theoretical Physics Group, LBNL, Berkeley, CA 94720, USA\\
\vspace{6pt}
{\it eggimon@lbl.gov}}
%\vspace{3pt}

$^3$ Physics Department, Stanford University, Stanford   CA 94305-4060, USA\\
\vspace{6pt}
 {\it kallosh@stanford.edu}

\vspace{10pt}

\underline{ABSTRACT}

\end{center}

We present explicit U-duality invariants for the $\mathbb{R}, \mathbb{C},
\mathbb{Q}, \mathbb{O}$  (real, complex,  quaternionic and octonionic) magic supergravities in four and five dimensions using
complex forms with a reality condition. From these invariants we derive an
explicit entropy function and corresponding stabilization equations which we
use to exhibit stationary multi-center $1/2$ BPS solutions of these $N=2$ $d=4$
theories, starting with the octonionic one with $E_{7(-25)}$ duality symmetry.
We generalize to stationary 1/8 BPS multicenter solutions of $N=8$, $d=4$
supergravity, using the consistent truncation to the quaternionic magic $N=2$
supergravity.    We present a general solution of non-BPS attractor equations of the STU truncation of  magic models. We
finish with a discussion of the BPS-non-BPS relations and
attractors in $N=2$ versus $N=5, 6, 8$.

\vfill

\end{titlepage}

\tableofcontents

\
%newpage

%%%%%%%%%%%%%%%

\section{Introduction}

The purpose of this paper is to present explicit  stationary multicenter
solutions of $N=8$,  $d=4$  supergravity \cite{Cremmer:1979up} and all  magic
$N=2$ supergravities \cite{Gunaydin:1983rk}. All these models are associated
with the Jordan algebras of $3\times 3$ Hermitian  matrices:
$J_3^{\mathbb{R}}$, $J_3^{\mathbb{C}}$, $J_3^{\mathbb{Q}}$, $J_3^{\mathbb{O}}$
and $J_3^{{\mathbb O}_s}$. Here $\mathbb{A}= \mathbb{R}, \mathbb{C},
\mathbb{Q}, \mathbb{O}$ are the four division algebras, with $\rm dim \;
\mathbb{A}= 1, 2, 4, 8$, while ${{\mathbb O}_s}$ is a the split form of
$\mathbb{O}$\ \cite{Ferrara:1997uz} with a quadratic norm invariant under
$O(4,4)$ (the indefinite signature means this is no longer a division algebra).
The octonionic magic $N=2$ model goes by the name exceptional since it is the
only one of the four magic $N=2$ supergravities which is not known to be a
consistent reduction of $N=8$: it's defining Jordan algebra $J_3^{\mathbb{O}}$
involves the real octonions with a quadratic norm invariant under $O(8)$
\cite{Ferrara:1997uz}, and is the only algebra mentioned above which one cannot
get from truncating $J_3^{{\mathbb O}_s}$.  Note that the connection to Jordan
algebras based on the division algebras defines a relation between the four
$N=2$ magic supergravities \cite{Gunaydin:1983rk} and the magic square
\cite{frt}. From a physics perspective, the main feature of the $N=8$ and the
magic N=2 supergravities which interests us is that for each case all the
vector fields, including the graviphoton, transform in a single irreducible
representation of U-duality group; these symmetries place strong constraints on
the entropy formula and the stabilization equations derived from it.

The black hole entropy formula for $N=2$ supergravities based on  symmetric
spaces is either known or can be established.  We extend this list to include
magic supergravities, for which we construct these entropies.  We can then give
a complete list of all symmetric spaces with their entropy formulas.

The duality symmetry of $N=8$, $d=4$  and $d=5$ supergravities is $E_{7(7)}$
and $E_{6(6)}$, respectively.  The duality symmetry of the  exceptional magic
$N=2$, $d=4$  and $d=5$ supergravities  is $E_{7(-25)}$  and $E_{6(-26)}$,
respectively. The number in the brackets stands for the difference between the
number of non-compact minus compact generators. For example,   in $E_{7(7)}$
and in $E_{7(-25)}$ the total number of non-compact  and compact   generators
of $E_7$ is the same, namely, $70+63=133$ and $54+79=133$. However, the
difference between  them is either $70-63=7$ for $E_{7(7)}$  or  $54-79=-25$
for $E_{7(-25)}$.

For the  octonionic $N=2$ supergravity we will find a new explicit entropy
function derived from the quartic invariant of the fundamental
$\mathbf{56}$-dimensional representation of $E_{7(-25)}$ and the relevant
``electric'' and ``magnetic'' cubic invariants of $E_{6(-26)}$.  This entropy
function then provide us with the most general explicit multicenter BPS
solution of the octonionic (exceptional magic) $N=2$ supergravity in terms of a
56-dimensional harmonic function with an arbitrary number of centers. We will
also present explicit entropy formulas for all the remaining magic
supergravities.

Recently, the generalization of the $N=2$ attractor equations in
\cite{Ferrara:1995ih}-\cite{FGK} was established for $N>2$  by a generalization
of the special geometry symplectic structure to all extended supergravities
\cite{Ferrara:2006em}. In particular, for $N=8$ supergravity a simple set of
algebraic equations was established which describe regular BPS and non-BPS
extremal black hole solutions.  In this later paper, a natural truncation of
the $N=8$ theory to it's largest consistent $N=2$ truncation, the quaternion
magic $N=2$ supergravity, made it's appearance.  To proceed with an explicit
demonstration of stationary multicenter 1/8 BPS solutions of $N=8$ $d=4$
supergravity we  use  this reduction to $N=2$. Hence, in addition to the $N=8$
model and the octonionic $N=2$ model, the N=2 magic supergravity based on the
Jordan algebra of quaternions, $J_3^{\mathbb{Q}}$, is of particular interest in
this paper.

We will argue that our new solution of quaternionic supergravity, depending on
a 32-dimensional harmonic function with any number of centers, are also the
most general 1/8 BPS stationary multicenter solution of $N=8$ $d=4$
supergravity, up to an overall $E_{7(7)}$ rotation.

We also will analyse multiple relations between BPS and non-BPS solutions in
$N=2$ and $N=5,6,8$ supergravities. It is known that the same bosonic solution
may be BPS or non-BPS depending on how it is embedded in a given supergravity
theory or how one identifies the bosonic vector fields with the specific
supersymmetric multiplets, see for example \cite{Khuri:1995xq} where many such
examples were given. In some $N=8$ examples in \cite{Khuri:1995xq} the relation
between BPS and non-BPS solutions uses the same embedding but requires a flip
of a sign from some of the charges. More recently such examples were discovered
in the context of non-supersymmetric attractors, see e. g.
\cite{Tripathy:2005qp}.

We will describe the general solutions of $N=2$ attractors and, in particular,
the general solution for non-BPS attractors of magic supergravities.

In summary, our paper is organized as follows.  In section 2 we summarize and
expand on known results for multi-center solutions in $N=2$ SUGRA with special
emphasis on the role of the a set of harmonic functions transforming in an
irreducible representation of the U-duality group.  Writing down the metric,
gauge and scalar fields depends crucially on a writing down a quartic
invariant, or an entropy function, constructed from this representation.  In
section 3 we describe two approaches for writing down the quartic invariants
for $N=8$ and the magic $N=2$ supergravities.  The first approach uses
connections to the five-dimensional U-duality group and makes novel use of a
set of complex matrices with a reality condition in expressing both the entropy
function and the attractor equations.  The second approach stresses the
connection of our quartic invariant to a set of invariants of the
six-dimensional U-duality group which allows us to use real matrices (the
quaternionic magic case is a pseudo-real exception).  In section 4 we develop,
as an example for the material in sections 2 and 3, the solutions for BPS
composites of octonionic magic $N=2$.  Section 5 follows with a description of
N=8 1/8 BPS composites in terms of a quaternionic $N=2$ sub-algebra.  In
section 6 we expand our circle of consideration to non-BPS extremal
multi-center solutions before closing with some final thought in section 7.

\section{A    description of
stationary multicenter solutions: \\
N=2 black hole composites}

%%%%%%%%%%%%%%%%%%%%%%%%%%%%%%%%%%%%%%%%%%%%%%%%%%

The exact BPS multicenter stationary black hole solutions in $N=2$ supergravity
have been worked out as a general case whenever there is a known expression for
the explicit single center black hole entropy, $S(p,q)$, as a function of
quantized charges\cite{Bates:2003vx}. The entropy formula
is given by the minimal value of the BPS black hole mass via
\be S(p,q) = \pi M^2(p,q; t, \bar t)|_{\rm attr}
= \pi I_1(p,q ),
\ee
where the moduli, $t(p,q), \bar t(p,q)$, are fixed near the black hole horizon
by the attractor mechanism \cite{Ferrara:1995ih}-\cite{FGK}.  There are three
invariants that the reader should be aware of in the context of black hole
attractors, it is important not to confuse them:
\begin{itemize}

\item $I_1((p,q; t, \bar t)$: This is the general symplectic invariant defined
in \cite{Ceresole:1995ca,FK}.  It can be written explicitly as:
\be
I_1(p,q; t, \bar t) = -{1\over 2} (p\;q)\, {\cal M}({\cal N}(t,\bar t))\,(p\;q)^T
= |Z(\Gamma)|^2 + |{\cal D}Z(\Gamma)|^2.
\ee
where ${\cal N}(t,\bar t)$ is the metric on the vector fields. Here $Z(\Gamma) = \langle \Gamma, \Omega \rangle$ and $\Gamma= (p^\Lambda ,q_\Lambda)$, $\Omega = (L^\Lambda, M_\Lambda)$.
This scalar quantity is an invariant of the group $OSp(2(n_v+1), \mathbb{R})$
which acts on both the charges and the symplectic sections.  It is manifestly
quadratic in the charges and non-negative.
Near the black hole horizon $N=2$ supersymmetry is restored and ${\cal D}Z=0$ which is equivalent to the requirement that
\be
\partial_t I_1(\Gamma; t, \bar t) = 0 \qquad \partial_{\bar t} I_1(\Gamma; t, \bar t) = 0
\ee
This equation defines the moduli near the horizon as functions of charges, $t(p,q), \bar t(p,q)$.

\item $I_1(p,q) = I_1(p,q; t(p,q), \bar t(p,q))$: This is the entropy function, $I_1(\Gamma)$,
derived from the function above restricted to $t$'s and $\bar t$'s
implicitly defined in terms of $p$'s and $q$'s through their attractor values.
This scalar is no longer invariant under $OSp(2(n_v+1), \mathbb{R})$ since
we have chosen particular values of $t$ and $\bar t$ for given $p$'s and $q$'s.
It is, however, invariant under the U-duality group if such exists, but it has
no particular polynomial properties as a function of the charges. The moduli near
the horizon are defined by equation \cite{FK}
\bea
 p^\Sigma +i {\partial I_1(p,q)\over \partial q_\Sigma} = 2i \bar Z(\Gamma) L^\Sigma \ , \qquad
 q_\Sigma -i {\partial I_1(p,q) \over \partial p^\Sigma}=  2i \bar Z(\Gamma) M_\Sigma  \label{old}\eea
which follows  from ${\cal D}Z(\Gamma)=0$ at the horizon. The holomorphic special coordinates $t^\Lambda= {X^\Lambda\over X^0}$  at the horizon follow by dividing eq. (\ref{old}) on the zero component of the same equation.
\be
t^{\Lambda}(p,q) = {p^{\Lambda} +i {\partial  I_1(\Gamma) \over \partial
q_{\Lambda}}\over p^0+i
{\partial  I_1 (\Gamma)\over \partial q_0}} \\
\label{stab1}\ee

\item $J_4(p,q)$: This is a quartic polynomial in the charges, invariant
under a U-duality group.  It appears in all the theories discussed in this
paper, it is
related to the entropy function via $I_1(p,q) = \sqrt{|J_4(p,q)|}$.
\end{itemize}

 Using the entropy function, the equations of motion for general black
hole solutions in $N=2$ supergravities were solved in
\cite{Bates:2003vx,Behrndt:1997ny}.  Examples of  non-static
multicenter solutions were given earlier in \cite{Bergshoeff:1996gg}
where it was also explained on the basis of these examples that the
most general expression for the metric can only depend on duality
invariants, the K\"{a}hler potential $K(X, \bar X)$ and the K\"{a}hler
connection ${\cal A}_\mu(X, \bar X)$. The general form  of multi-center BPS black hole solutions was presented in \cite{Behrndt:1997ny}.

The  explicit BPS multicenter stationary black hole solutions in $N=2$
supergravity were found in \cite{Bates:2003vx,Behrndt:1997ny}. One can solve
the 1/2 BPS equations by introducing a harmonic symplectic doublet:
\be
\H(\vec x)= (H^{\Lambda}, H_\Lambda) = {\bf h}+ \sum_{s=1}^{n}{{\bf \Gamma}_s
\over |\vec x-\vec x_s|}.
\ee
From BPS equations one can prove that this doublet is proportional to the imaginary part of the covariantly holomorphic symplectic section $\Omega = (L^\Lambda, \, M_\Lambda )$:
\be
\H(\vec x)= 2 {\rm Im} \, \langle {\bf H}(\vec x), \overline \Omega \rangle \, \Omega \equiv i (H^\Lambda \overline M_\Lambda - H_\Lambda \overline L^\Lambda) \Omega \ , \qquad \langle \Omega , \overline \Omega \rangle = -i
\ee

We will refer to ${\bf h}$ and $\G$ as {\em fundamentals} since for the
theories which appear in our paper they transform as fundamentals under the
$E_7$ U-duality group. A symplectic invariant $I_1(\Gamma, t, \bar t)$ is now replaced by the $\vec x$-dependent symplectic invariant $I_1(\H(\vec x), t, \bar t)$. The other basic ingredient is the symplectic pairing of two charges which
induces one on the harmonic functions:
$$
 \langle \G_1,\G_2 \rangle  = p_1^\Lambda q_{2\,\Lambda} - p_2^\Lambda q_{1\,\Lambda}
 \;\hookrightarrow\;  \langle \H_1,\H_2 \rangle.
$$
 Once equipped with these structures,  one can prove that  the BPS equations require that
 \be
\langle H,{\cal D} \Omega \rangle =0
 \ee
 which results in
\be
\partial_t I_1(\H(\vec x); t, \bar t) = 0 \qquad \partial_{\bar t} I_1(\H(\vec x); t, \bar t) = 0
\ee
This is the stabilization equation for the symplectic invariant  which can also be translated into a relation
\be
I_1((\H(\vec x); t, \bar t)|_{\langle \H(\vec x), {\cal D} \Omega \rangle  = 0}= I_1((\H(\vec x))
\ee
This leads to
\bea
 H^\Sigma +i {\partial I_1(\H)\over \partial H_\Sigma} = 2i \bar Z(\H)  L^\Sigma \ , \qquad
 H_\Sigma -i {\partial I_1(\H) \over \partial H^\Sigma}=  2i \bar Z(\H) M_\Sigma  \label{new}\eea
The special coordinates $t^\Lambda= {X^\Lambda\over X^0}$ solve the equations
of motion, stabilization equations of the same form as the attractor equations
for moduli near the horizon, and are given by
\be
t^{\Lambda}(\vec x) = {H^{\Lambda} +i {\partial  I_1(H) \over \partial
H_{\Lambda}}\over H^0+i
{\partial  I_1 (H)\over \partial H_0}} \\
\label{stab2}\ee
The stationary metric is
\be
ds_4^2= - I_1^{-1}(\vec x) (dt+\vec \omega d \vec x)^2+ I_1(\vec x)
d\vec x^2,\qquad \nabla \times \vec  \omega= \langle \H, \nabla \H \rangle.
\ee
Note that $e^{- 2 K(X, \bar X)}$ may or may not be equal to $ I_1(\vec x)$,
depending on the choice of the  K\"{a}hler gauge. The symplectic invariant $
I_1$ is  invariant under K\"{a}hler transformations,  $K(t, \bar t) \rightarrow
K(t, \bar t)+ f(t) + \bar f(\bar t)$, whereas the K\"{a}hler potential can be
changed by a sum of the holomorphic and anti-holomorphic function. In
\cite{Behrndt:1997ny} an ansatz was used with $K= -2U$ and $H= i \Omega_0 -
i\overline \Omega_0$, where $\Omega_0= (X^\Lambda,  F_{\Lambda})$ is the
holomorphic section. This section also transforms under  K\"{a}hler
transformations, $\Omega_0 \rightarrow \Omega_0 e^{-f(t)}$. In
\cite{Bates:2003vx} the ansatz for the harmonic doublet is $H= i e^{-U+K/2}
(e^{ -i\alpha} \Omega_0 - e^{ i\alpha}\overline \Omega_0)$ where $\alpha$ is
the argument of the central charge. We prefer to codify the solutions by the
choice of the harmonic function at infinity, ${\bf h}$.  Using this boundary
value, we can give the values of the special coordinates,  the metric and
vector field strength and do not have to define combinations which are not
invariant under K\"{a}hler transformations. For example, the relation between
our ${\bf h}$ and the quantities in \cite{Bates:2003vx} is ${\bf h}=
(e^{-i\alpha} e^{K/2}\Omega_0)_{\infty}$. In fact, only the product of all 3
terms is invariant under K\"{a}hler transformations. Therefore we can simply
codify our solution with arbitrary ${\bf h}$. Note also that in our equation
(\ref{new}) the analogous term $\bar Z(\H)  L^\Sigma$ is K\"{a}hler invariant
since the transformations on $\bar Z(\H)$ and on $L$ cancel and we do not have
to identify these terms separately.

The integrability condition for the multi-center solution is
\be
\langle \H, \triangle \H \rangle=0 \qquad \Leftrightarrow \qquad \langle
\H(\vec x_s), {\bf\Gamma}_s \rangle =0 \qquad \Leftrightarrow \qquad
\sum_{t=1}^{n}{\langle {\bf\Gamma}_s, {\bf\Gamma}_t \rangle  \over |\vec
x_s-\vec x_t|}+ \langle  {\bf\Gamma}_s, {\bf h} \rangle=0
\label{integr}\ee
Hence when our generic set of $n+1$ fundamentals, $(\G_s,{\bf h})$, satisfies
the integrability conditions, we constrain the distances between the $\Gamma_s$
fundamentals, $|\vec x_s-\vec x_t|$.  Only when two of these fundamental charge
vectors are mutually local is their relative distance unconstrained.  For the
case when all the fundamentals are mutually local there are no constraints on
the distances; they are just moduli.

   The fundamental which sets the asymptotic behavior, ${\bf h}$, also must
satisfy some conditions if the metric is to have the correct normalization and
the integrability conditions are to be solved:
\be
I_1({\bf h})=1,\qquad <{\bf h},\G> =0,\qquad (\G = \sum_s \G_s)
\ee
The vector fields in electric basis, using a mixture of spherical coordinates
$(r_s, \theta_s, \phi_s)$ around each center $\vec x_s$,  are:
\be
\label{gaugefields}
A_\Lambda= \partial_{H^{\Lambda}} \Big (\ln I_1(H)\Big ) (dt+\omega)
- \sum_s \cos \theta_s \, d\phi_s \otimes \Gamma_{\Lambda}.
\ee
The features of the geometry at very large $|\vec x|\gg |\vec x_s-\vec x_t|$
can be read off from the asymptotic form of the solution.  In this
approximation
\be
\H(\vec x) \approx  {\bf h} +{{\bf \Gamma} \over |\vec x|}\ ,  \qquad {\bf
\Gamma}= \sum_{s=1}^{n}{\bf \Gamma}_s
\ee
and
\be
\label{omegaexp}
 \vec \omega\approx \sum_{s<t} \langle {\bf \Gamma}_s, {\bf \Gamma}_t
 \rangle \;
 \vec e_{st} \times {\vec x\over |\vec x|} \ , \qquad \vec e_{st}= {\vec x_s-\vec x_t \over |\vec x_s-\vec x_t|}
\ee
Far from the core, this looks like a spherically symmetric black hole (see
\cite{gimonetal} for more details):
\be
ds_4^2= - e^{2U} dt^2+ e^{-2U} d\vec x^2 \ , \qquad e^{-2U(\vec x)}= I_1\Big
({\bf h}+{{\bf\Gamma} \over |\vec x|}\Big )
\label{static}\ee
whose entropy is equal to
\be
 S({\bf\Gamma})= \pi{I_1({\bf\Gamma})}.
\ee
Note that any measured angular momentum derived from our expansion of $\omega$
in (\ref{omegaexp}) comes from terms which are order $1/|\vec x|^2$ in the
metric and should be matched with a measurement at a similar order for the
dipole moments of the gauge fields in (\ref{gaugefields}).  Thus there is no
contradiction with our knowledge that simple charged supersymmetric black holes
have no angular momentum; if you measure angular momentum you also measure
dipoles.

To recap, we would like to stress the special features of these $N=2$
solutions. For a solution with $n$ centers we needed $n+1$ constant fundamentals
(symplectic doublets) :
\be
{\bf h} = (h^\Lambda , h_\lambda) \ , \qquad \G_s = (p^\Lambda, q_\Lambda) _s
\qquad s=1, \dots , n.
\ee
giving these important qualities to our solutions:
\begin{itemize}
  \item {\it An attractor at infinity}:  This means that
the values of all our special coordinates at infinity are specified by the
values of the 1st fundamental, ${\bf h}$ and are completely independent of all
the other fundamentals,  $\G_s$. This follows from the limit of eq.
(\ref{stab2}) at infinity.
\be
t^{\Lambda}(\vec x)|_{|\vec x|\rightarrow \infty} \quad \rightarrow \quad
{h^{\Lambda} +i {\partial  I_1(h) \over \partial h_{\Lambda}}\over h^0+i {\partial  I_1 (h)\over \partial h_0}} \\
\label{stabinf}\ee

  \item {\it An attractor at each center}: this is a familiar feature from
BPS 1-center black holes where the values of the moduli near the black hole
horizon are independent on the values of moduli at infinity. In our case, with
many centers, the values of special coordinates at each center are independent
of all fundamentals except the one at the given center. The limit of eq.
(\ref{stab2}) near each center is given by
\be
t^{\Lambda}(\vec x)|_{|\vec x|\rightarrow |\vec x_s|} \quad \rightarrow \quad
{\Gamma_s^{\Lambda} +i {\partial  I_1(\Gamma_s) \over \partial \Gamma_{s \Lambda}}\over \Gamma_s^0+i {\partial  I_1 (\Gamma_s)\over \partial \Gamma_{s 0}}} \label{stabs}\ee
\end{itemize}
Apart from the nice clarity of the attractor behavior above, at each center
and at infinity, writing our
solution in terms of $n+1$ fundamentals of our duality group yields two more
simple features not yet discussed in the literature:
\begin{itemize}
    \item  The constant term in our integrability condition, (\ref{integr}), at
each center
is simply written in terms of the symplectic pairing between ${\bf h}$ and
${\bf \G_s}$ without any reference to the phase of the central charge such as
in \cite{Behrndt:1997ny,Bates:2003vx} (this simplification is also observed in
\cite{gimonetal}).  This makes the construction, in principle, easier to extend
to a non-BPS form.
    \item The mass of the black hole in terms of the charges and asymptotic moduli
can be written very compactly in terms of ${\bf h}$ and $\G$ by expanding
$I_1(\vec x)$ to leading order in $1/|\vec x|$.
\end{itemize}

In what follows we will find that for most of the magic supergravities cases it
is most convenient to generalize the form of the black hole composite solutions
above: magic symmetries are best manifested using a double index, i.e matrix,
notation instead of a single index $\Lambda$ in coordinates and charges.  We
write real quantities, such as charges, as constrained complex matrices with a
reality condition of type, $t^*=\Omega\, t\, \Omega^T$, and holomorphic
quantities using the associated notion of complex conjugation.

\section{Black hole entropies of  magic $N=2$ and of $N=8$ supergravities}

%%%%%%%%%%%%%%%%%%%%%%%%%%%%%%%%%%%%%%%%%%%%%%%%%%%%

In order to find black hole composite solutions in four dimensions we require
an explicit entropy formula in terms of generic black hole charges. Such
formulas are known for {\it symmetric spaces}; the appropriate $N=2$
supergravity coupled to vector multiplets are classified in
\cite{Cremmer:1984hc}-\cite{deWit:1992wf}. These models include the reducible
spaces $\Big [{SU(1,1)\over U(1)}\Big]^3$, ${SU(1,1)\over U(1)}\times {SO(P+2,
2)\over SO(P+2)\times SO(2)}$, and the complex projective space  ${SU(1,
n)\over SO(n)\times U(1)}$. In Table 2 of \cite{deWit:1992wf}  the first two
cases are $L(0,0)$ and $L(0,P)$ respectively. For these  three classes of $N=2$
$d=4$ supergravities  the entropy formula is known for generic set  of charges
since the attractor equations  \cite{Ferrara:1995ih}-\cite{FGK} have been
solved for these three spaces in \cite{Behrndt:1996hu}-\cite{Behrndt:1997fq}, respectively. The remaining symmetric spaces for which
the entropy formulas have not been established so far are connected to Jordan
algebras, according to \cite{Cremmer:1984hc}. These are the magic
supergravities.

We will present here the new $d=4$ entropy formulas for magic $N=2$
supergravities using the relation between the $d=4$,  $d=5$ and $d=6$ dualities
\footnote{Recently the $d=4\;  \Leftrightarrow \; d=5$ relation between
particular black hole solutions has been used  in counting of the BPS black
hole degeneracy \cite{Shih:2005qf}.}. We will also give some new forms of the
entropy formula for $N=8$ supergravity which can be easily compared with the
entropy of the octonionic $N=2$.

\subsection{ Universal $d=4\;  \Leftrightarrow \; d=5$ relation}

Consider  the  splitting
\be
G_4 \rightarrow G_5 \times SO(1,1)
\label{decomp}\ee
where $G_5$ is the duality group in $d=5$  and $G_4$ is the duality group in
$d=4$. The quartic invariant of the relevant duality groups $G_4$ can be
constructed using the cubic invariant of $G_5$. In all cases we have to split
the total set of electric and magnetic charges into the zero component and the
rest,
\be
(p,q)= (p^0, p^{I}; q_0, q_{I})\ , \qquad I= 1, \dots , n_v
\ee
This splitting is of the type
\be
\mathbf{R}_{p,q}= \mathbf{R}_p +\mathbf{1}_p + \mathbf{R}'_q +\mathbf{1}'_q
\ee
where $\mathbf{R}_p$ is the five-dimensional representation and $\mathbf{R}'_q$
is its contravariant representation:
\bea
\mathbf{R}_p &\rightarrow& (p^I) \qquad \mathbf{R}'_q \rightarrow (q_I)\\
\mathbf{1}_p &\rightarrow& (p^0)\qquad \mathbf{1}'_q \rightarrow (q_0)
\eea
All of these are real representations of $G_4$ or $G_5$ as appropriate. In the
magic models $\mathbf{R}_{p,q}$ refers to the following representations:
\bea
J_3^{\mathbb{O}}&&\qquad \rightarrow \qquad \mathbf{56} \qquad  \rm {  fundamental \; of}
\; E_{7(-25)}\nonumber \\
J_3^{\mathbb{Q}}&&\qquad \rightarrow \qquad \mathbf{32} \qquad  \rm { chiral\;  spinor \; of} \;
SO^*(12)\\
J_3^{\mathbb{C}}&&\qquad \rightarrow \qquad \mathbf{20} \qquad  \rm {  threefold \; antisymmetric \;
(selfdual)  \; of} \; SU(3,3)\nonumber \\
J_3^{\mathbb{R}}&&\qquad \rightarrow \qquad \mathbf{14}' \qquad \rm { threefold
\; antisymmetric \; traceless  \; of} \; Sp(6,\mathbb{R}) \nonumber
\eea
The general formula for the quartic invariant of the four-dimensional duality
groups $G_4$ related to its $d=5$ cubic invariants is:
\be
J_4 (p^0, q_0, p^I, q_I) = -(p\cdot q)^2 + 4\Big (q_0 I_3(p) - p^0 I_3(q) +\{
I_3(q), I_3(p)\}   \Big) \ .
\label{quartic} \ee
 where the cubic invariants of the five-dimensional duality groups $G_5$ are given by
\be
I_3(p) = {1\over 3!} d_{IJK} p^I p^J p^K \ ,  \qquad I_3(q) = {1\over 3!}
d^{IJK} q_I q_J q_K
\label{cubic}\ee
and the scalar product of charges and the Poisson bracket of cubic invariants
are defined as follows
\be
p\cdot q \equiv p^0 q_0 + p^I q_I \ , \qquad \{ I_3(q), I_3(p)\}\equiv
{\partial I_3(q)\over \partial q_I} \; {\partial I_3(p)\over \partial p^I}
\ee
This structure of quartic invariants in terms of the cubic ones is valid for
all  cubic systems, where both $d_{IJK}$ as well as $d^{IJK}$ are known, in
particular,  it can be used for $N=8$ and for all the magic $N=2$ theories.
Each model corresponds to specific values for $d_{IJK}$ and $d^{IJK}$ in
equation (\ref{cubic}) which form the cubic invariants of $G_5$. The quartic
formula above can be derived following \cite{Ferrara:1997uz,Pioline:2005vi}
%\cite{Pestun:2005ni}
using the Freudenthal triple system  that applies for any decomposition of the
type given in eq.(\ref{decomp}). The properties of the coset spaces which one
encounters in magical supergravities can be inferred from
\cite{Helgason,Gilmore}.

For $N=8$ the classic example is $E_{7(7)}\rightarrow E_{6(6)}\times SO(1,1)$.
The quartic invariant of the $E_{7(7)}$ and the cubic invariant of the
$E_{6(6)}$ have been in the literature a long time: they were used for the
single black hole entropy in $N=8$, $d=4$,  \cite{Kallosh:1996uy,Ferraraetal}
and $N=8$, $d=5$, \cite{FK2}, respectively.  In this section, we will exploit
the relationship between $N=8$ and the octonionic magic $N=2$ to efficiently
present invariants for all the magic supergravities.  We can use these, in
turn, to write explicit black hole composite solutions.

Let us quickly review the magic $N=2$ supergravities. These are labeled using
the four division algebras $\mathbb{A}$: $\rm dim \mathbb{A}$ is $1$ for real
numbers $\mathbb{R}$, $2$ for complex numbers $\mathbb{C}$, $4$ for quaternions
$\mathbb{H}$, and $8$ for octonions $\mathbb{O}$.  $N=8$ supergravity
corresponds to the split octonions, which do not form a division algebra.

The scalars in $N=8$ supergravity are in the cosets ${G_5\over
H_5}={E_{6(6)}\over Usp(8)}$ and ${G_4\over H_4}={E_{7(7)}\over SU(8)}$   in
$d=5$ and $d=4$, respectively.  In each case, the vector fields are in the real
representations ${\bf 27}$ of $E_{6(6)}$ and ${\bf 56}$ of $E_{7(7)}$.

For magic supergravities the cosets in $d=5$ are, in the order of increasing
dimension of the division algebra $\rm dim \mathbb{A}$:
\be
d=5: \qquad {G_5\over H_5}\Rightarrow  \qquad {SL(3;\mathbb{R})\over SO(3)}\ ,
\qquad {SL(3;\mathbb{C})\over SU(3)}\ , \qquad {SU^*(6)\over Usp(6)}\ , \qquad
{E_{6(-26)}\over F_4}
\label{G0}\ee
and in four-dimensional theories
\be
d=4: \;\; {G_4\over H_4}\Rightarrow \;\; {Sp(6;\mathbb{R})\over SU(3)\times
U(1)}\ , \;{SU(3, 3)\over SU(3)\times SU(3)\times U(1)}\ , \; {SO^*(12)\over
SU(6)\times U(1)}\ , \; {E_{7(-25)}\over E_6\times U(1)}
\ee
The number of vector fields in the five-dimensional versions of magic
supergravities is
\be
 d=5\ ,  \qquad n_v= 3({\rm dim} \mathbb{A}+1) \, \Rightarrow \qquad n_v= 6, 9, 15, 27
   \qquad \rm for \quad \mathbb{R}, \mathbb{C}, \mathbb{H}, \mathbb{O},
\ee
with one of the vectors coming from the gravity multiplet and transforming as a
singlet of $H_5$. The ``magic" comes from representing the charges for all
these vector fields as real parameters of a self-adjoint 3x3 matrix over the
appropriate algebra (elements of the Jordan algebra $J_3^{\mathbb{A}}$). In
$d=4$, Kaluza-Klein reduction adds one more vector field, and so the numbers
are:
\be
 d=4 \ , \qquad \tilde n_v= 3({\rm dim} \mathbb{A}+1)+1 \, \Rightarrow
 \qquad \tilde n_v= 7, 10, 16, 28   \qquad \rm
 for \quad \mathbb{R}, \mathbb{C}, \mathbb{H}, \mathbb{O},
\ee
with one of the vectors coming from the gravity multiplet and transforming as a
singlet of $H_4$.

We will demonstrate two complementary methods for establishing the
explicit form of the quartic/cubic invariants and related d=4,5 black hole entropies.
\begin{enumerate}

\item Starting in five-dimensions, we use the {\it manifest  five-dimensional duality}
group $G_5$ for all of our models, with  $E_{6(6)}, E_{6(-26)}, SU^*(6),
SL(3;\mathbb{C}), SL(3;\mathbb{R})$-symmetry, and the simple features of the
corresponding Jordan algebras to write down the appropriate cubic invariants.
Operationally, this entails representing the real coordinates of the N=8 theory
and the very special real geometry of the $N=2$ theories in terms of complex
anti-symmetric matrices obeying a reality condition.  We then obtain the
holomorphic coordinates of the corresponding four dimensional scalar manifold
geometry by relaxing the reality condition. To the reader, this may seem
different from the well known case where one starts with real matrices in $d=5$
complexify them term by term to get the holomorphic coordinates in $d=4$. In
fact, the difference is cosmetic rather than structural, due to the convenience
of using a double index notation; we will make this clear in an explicit
mathematical way.

  \item We rearrange our charges and coordinates using just $G_6$ manifestly-symmetric
representations.  In this case we have to further split the five-dimensional duality symmetry
down to the six-dimensional one:
\be
G_5 \quad \Rightarrow \quad G_6\times SO(1,1)
\ee
The duality group, $G_6$, for magic supergravities is $SO(1, \rm dim
\mathbb{A}+1)$, similar to the duality group  $SO(5,5)$ for the $N=8$ model.
For a $IIB$ oxidation, tensor charges (self-dual strings) will appear in vector
representations of $G_6$ and vector charges (point charges and their magnetic
duals) will appear as real spinors. The advantage to this method is that, in
all but one case,  we can reconstruct the cubic invariant in d=5 using just the
metric and gamma matrices of $G_6$.  For the quaternionic
magic the eight dimensional real representation of $SO(1,5)$ is really
the combination of 2 pseudo-real spinors of the same chirality;  the answer in
this case is only slightly more complicated.  In the first cases ($\rm \dim
\mathbb{A} \ne 4$), the d=4 holomorphic coordinates are a simple
complexification of the d=6 real coordinates. The details for the more
subtle quaternionic case are presented in the Appendix B.

While overall not as manifestly symmetric, the six-dimensional approach
provides us with more than a cross-check for our first method.  The choice of
splitting for $G_5 \to G_6 \times SO(1,1)$ also yields some valuable insight
into the oxidation process, e.g. the split into tensor and vector multiplets
above.

\end{enumerate}

\subsection{ $G_5$ manifestly-symmetric entropy of $d=4$ BPS black holes}

Our goal at this point is to develop a simple expression for the cubic
invariants of the five-dimensional magic supergravity; using these we can then
extract a $G_5$ manifestly-symmetric formula for the quartic invariant in four
dimensions using the formula in (\ref{quartic}).  A very efficient way to
accomplish our task is to use the relationship of these theories to Jordan
algebras and the N=8 maximally supersymmetric theory.

Let us demonstrate with the familiar case of $N=8$ supergravity
\be
E_{7(7)} \quad \rightarrow \quad E_{6(6)}\times SO(1,1)
\ee
and the octonionic magic $N=2$ supergravity with
\be
E_{7(-25)} \quad \rightarrow \quad E_{6(-26)}\times SO(1,1).
\ee
The respective five-dimensional charge spaces for these theories correspond to
the Jordan algebras $J^{{\mathbb O}_s}_3$ and $J^{{\mathbb O}}_3$.  We can
represent the ${\bf 27}$ elements for both algebras conveniently using the
(faithful) traceless anti-symmetric representations of $Usp(8)$ and $Usp(6,2)$.
This is similar in spirit to the use of $SU(8)$ and $SU(6,2)$ to construct
quartic invariants in \cite{Gunaydin:2004md}, but our results for the cubic
invariants of $E_{6(6)}$ and $E_{6(-26)}$ do not descend straightforwardly from
these quartic invariants.  It would be interesting to have a better
understanding of the connection between the two.

We will now quickly review the details necessary to build these
representations. We start by defining \cite{Helgason,Gilmore} the $(8\times 8)$
symplectic matrix $\Omega$ and a metric preserved by $Usp(8)$ and $Usp(6,2)$
\be
\Omega = {\mathbb I}_4 \otimes \begin{pmatrix}
  0 & -1\\
 1 & 0
\end{pmatrix},\qquad
K^{(a,b)}= \alpha \otimes
\begin{pmatrix}
  1 & 0 \\
0 & 1
\end{pmatrix}
\label {K}\ee
which will allow us to treat both cases, $E_{6(6)}$ in  $Usp(8)$ basis and
$E_{6(-26)}$ in $Usp(6,2)$  basis, simultaneously.  The $(4\times 4)$ matrix
$\alpha$ is
\be
\alpha =\begin{pmatrix}
  1 & 0 & 0 & 0 \\
  0 & 1 & 0 & 0 \\
  0 & 0 & 1 & 0 \\
  0 & 0 & 0 & 1
\end{pmatrix} \quad { \rm for }\quad   Usp(8)\ ,  \qquad \alpha =\begin{pmatrix}
  -1 & 0 & 0 & 0 \\
  0 & -1 & 0 & 0 \\
  0 & 0 & -1 & 0 \\
  0 & 0 & 0 & 1
  \end{pmatrix} \quad { \rm for } \quad Usp(6,2),
\label{alpha}\ee
and $\alpha=\alpha^t$, $\alpha^2=1$.  We will label the two different metrics
either $K^{(0,4)}$ or $K^{(3,1)}$.

We are now ready to exhibit the {\em real} representation $\mathbf{27}$ of
$E_{6(6)},\, E_{6(-26)}$ taken in the $Usp(8),\, Usp(6,2) $ basis by using the
map $t^I \rightarrow t^{ab}$ where $t^{ab}$ is the antisymmetric traceless
matrix, $t^{ab}=-t^{ba}$, $Tr \,t \equiv  t^{ab} \Omega_{ab}=0$ satisfying a
reality condition
\be
 t^*_{ab}= \tilde \Omega_{aa'}\tilde \Omega_{bb'} t^{a'b'}
\qquad \Leftrightarrow \qquad t^*= \tilde \Omega\, t \, \tilde \Omega^T \ ,
\qquad
\label{reality}\ee
where
\be
 \tilde \Omega = \begin{pmatrix}
  0 & -\alpha \\
\alpha & 0
\end{pmatrix}= K \Omega = \begin{pmatrix}
 \alpha & 0 \\
0 & \alpha
\end{pmatrix} \begin{pmatrix}
  0 & -I \\
 I & 0
\end{pmatrix} \ , \qquad  \Omega = \begin{pmatrix}
  0 & -I \\
I & 0
\end{pmatrix}
\label{tilde}\ee
With our framework, we quickly write down our cubic invariant as:
\be
d_{IJK}t^I t^J t^K = Tr( \Omega \, t\,  \Omega \, t\,  \Omega \, t )
\ee
which is a real polynomial. For the contragradient representation  we can take
\be
d^{IJK}t'_I t'_J t'_K= Tr( \Omega' \, t'\,  \Omega' \, t'\, \Omega' \, t' )\ ,
\qquad t'= t'_{ab} \ , \qquad  \Omega'= \Omega^{\mathbf{T}}.
\ee

Using these representations, the cubic invariants of $E_{6(6)}$ and $E_{6(-26)}$
in $Usp(8)$ and $Usp(6,2)$  basis, respectively, are given by
\be
\label{cubic1}
I_3(p)= \frac1{3!}\, Tr( \Omega \, p\,  \Omega \, p\,   \Omega \, p )\ , \qquad
I_3(q)= \frac1{3!}\, Tr( \Omega' \, q\,  \Omega' \, q\,   \Omega' \, q )
\ee
Here
\be
p^* = \tilde \Omega\, p \, \tilde \Omega^T \ , \qquad Tr\,( p\Omega) \equiv
p^{ab}\Omega_{ab}=0 \ ,   \qquad  q^* = \tilde \Omega\, q \, \tilde \Omega^T \
, \qquad Tr\,( q \Omega)=0\ .
\ee
The reality and trace condition give exactly 27 real entries for the complex
matrices $p^{ab}$ and $q_{ab}$; the complex two index notation is simply a
convenient way to represent the real numbers $p^I$ and $q_I$.  The only
difference between the the cubic invariants of $E_{6(6)}$ and $E_{6(-26)}$
comes from the reality condition, driven by a different choice of
$\tilde\Omega$ coming from different versions of the metric $K$.  The the fact
the cubic invariants are not explicitly different in terms of complex matrices,
but only different through their respective reality constraint is a direct
consequence of the fact that $E_{6(6)}$ and $E_{6(-26)}$ are different real
forms of the same complex $E_6$ algebra.

The reality constraint (\ref{reality}) and the tracelessness condition $ t^{ab}
\Omega_{ab}=0$ can be solved and one finds that
\be
{\rm Re}\,  t = \begin{pmatrix}
 A_1 & S \\
-\alpha S \alpha  & \alpha A_1 \alpha
\end{pmatrix} \ , \qquad  {\rm Im}\,  t= \begin{pmatrix}
 A_2 & A \\
\alpha A \alpha  & -\alpha A_2 \alpha
\end{pmatrix}
\ee
where $S^t= \alpha S \alpha$, $Tr\, S=0$ and $A^t= -\alpha A \alpha$. This
means that ${\rm Re}\,  t$ depends on $n^2-1=15$ real entries since $A_1$ and
$A_2$ and $A\alpha$ are antisymmetric,  and $S\alpha$  is symmetric.    ${\rm
Im}\,  t$ depends on $n^2-n= 12$ real entries. Together $t$ has 27 real
entries. The same forms apply for the electric and magnetic charges, in
agreement with $n_v= 3({\rm dim} \mathbb{A}+1)= 3(8+1)$ for octonions.

If we want to use only real entries for magnetic and electric charges, where
$p={\rm Re}\, p +i {\rm Im}\,  p = p_1+ip_2$, we can use
\be
\label{preal}
p^* = \tilde \Omega\, p \, \tilde \Omega^T  \quad \Rightarrow \quad   p_1=
\tilde \Omega   p_1 \tilde \Omega^T \ , \qquad   p_2= - \tilde \Omega   p_2
\tilde \Omega^T
\ee
The cubic invariants in terms of the real entries  are
\be
\label{pcubicreal}
I_3(p)= \frac1{3!}\,Tr( \Omega \, p\,  \Omega \, p\,   \Omega \, p )=
\frac1{3!}\,Tr( \Omega \, p_1 \,  \Omega \,   p_1 \,   \Omega \,   p_1 )-
\frac12\, Tr( \Omega \,   p_1 \, \Omega \, p_2 \,   \Omega \,   p_2 )
\ee
The electric charges are in the contragradient representation, and so we expand
$q$ as $q ={\rm Re}\, q  - i (-{\rm Im}\,q)  = q_1 - i q_2$, with the analogous
equations to eqs. (\ref{preal}) and (\ref{pcubicreal}).

{\it Three other $G_5$-invariant magic entropies }

Consider the {\it quaternionic magic}  case with $G_5 = SU^*(6),\, H_5 =
Usp(6)$.  Our charges are in the ${\bf 15}$, the two-fold anti-symmetric
representation $SU^*(6)$, which decomposes as the ${\bf 14}$ traceless
anti-symmetric of $Usp(6)$ plus a singlet.  This anti-symmetric representation
of $Usp(6)$ is not the truncation to the common $6\times 6$ matrix of the two
types of $8\times 8$ matrices above (as opposed to what happens in
\cite{Gunaydin:2004md}).

 We write the quaternionic magic charges using the matrix $U^{\Lambda \Sigma},
\,U=-U^t$ with the reality condition $U^*= \Omega U\Omega^T$, giving us $n_v=
3({\rm dim} \mathbb{A}+1)= 3(4+1)= {6\cdot 5\over 2}$, so we that we end up
with 15 electric and 15 magnetic charges.  The real cubic forms are:
\be
{1\over 3!} d_{IJK}t^I t^J t^K =  {1\over 3!\cdot 2^3} \epsilon_{\Lambda \Sigma
\Delta \Gamma \Pi \Omega} U^{\Lambda \Sigma} U^{\Delta\Gamma} U^{\Pi
\Omega}\equiv {\rm Pf}\,  U
\ee
\bea
I_3(p) &=& {1\over 3!\cdot 2^3} \epsilon_{\Lambda \Sigma \Delta \Gamma \Pi
\Omega}
p^{\Lambda \Sigma} p^{\Delta\Gamma} p^{\Pi \Omega}\equiv +{\rm Pf}\,  p \\
I_3(q) &=& {1\over 3!\cdot 2^3} \epsilon^{\Lambda \Sigma \Delta \Gamma \Pi
\Omega} q_{\Lambda \Sigma} q_{\Delta\Gamma} q_{\Pi \Omega}\equiv -\,{\rm Pf}\,
q
\eea
In terms of real  charges $p_1= \Omega\, p_1 \Omega^T$ and $p_2= -\Omega\, p_2
\Omega^T$ where $p^*= \Omega\, p \Omega^T$ and $p=p_1+i p_2$ we have
\bea
I_3(p_1, p_2) = {1\over 3!\cdot 2^3} \epsilon_{\Lambda \Sigma \Delta \Gamma \Pi
\Omega}\, p_1^{\Lambda \Sigma}( p_1^{\Delta\Gamma} p_1^{\Pi \Omega}- 3
p_2^{\Delta\Gamma} p_2^{\Pi \Omega})
\eea
and analogous for $q = q_1 -i q_2$.
 When we use these expressions (cf. section 5) to obtain a quartic invariant, we
get an expression which matches exactly with the quartic invariant for
quaternionic magic in \cite{Gunaydin:2004md}.  This quartic invariant also
appears explicitly in \cite{Andrianopoli:1997pn} within the context of $N=6$
supergravity.

The next case is {\it complex magic} with $G_5/H_5 =  SL(3,\mathbb{C})/SU(3)$
and $n_v= 3(2+1)= 9$. We truncate $U^{\Lambda\Sigma}$ by writing it in terms of
a symmetric 3x3 matrix $U^{I\bar I}$:
\be
U^{\Lambda\Sigma} \to U^{I\bar I} ,\qquad U_{\Lambda\Sigma} \to U_{I\bar I}.
\ee
The reality condition makes $U_{I\bar I}$  hermitian (the trace is an invariant
of $H_5$).  The cubic invariants in term of real charges, where $p^{I\bar I} =
{\rm Re}\,  p^{I\bar I} + i {\rm Im}\, p^{I\bar I}= p_1^{I\bar I} + i
p_2^{I\bar I}$ and $q_{I\bar I} = {\rm Re}\,  q_{I\bar I} - i (-{\rm Im}\,
q_{I\bar I})= q_{1\, I\bar I} - i \,q_{2\, I\bar I}$ are:
\bea
I_3(p) &=& {1\over 3!}\epsilon _{IJK}
\epsilon_{\bar I \bar J \bar K} \,p^{I\bar I}p^{J\bar J} p^{K\bar K}=  \det p\\
&=& {1\over 3!} \epsilon _{IJK} \epsilon_{\bar I \bar J \bar K} \,p_1^{I\bar I}
(p_1^{J\bar J} p_1^{K\bar K}- 3 p_2^{J\bar J} p_2^{K\bar K}) \nonumber\\
\nonumber\\
I_3(q) &=& {1\over 3!} \epsilon ^{IJK} \epsilon^{\bar I \bar J \bar K} q_{I\bar
I}q_{J\bar J} q_{K\bar K}= \det q \\
&=& {1\over 3!} \epsilon ^{IJK} \epsilon^{\bar I \bar J \bar K} \,q_{1\, I\bar
I} (q_{1\, J\bar J} q_{1\, K\bar K}- 3 q_{2\, J\bar J} q_{2\, K\bar K}).
\nonumber
\eea
The last case, {\it real magic} has $t^I$  in $\mathbf{6}$ of
$SL(3,\mathbb{R})$, and $n_v=3(1+1)= 6$.
\be
{1\over 3!}d_{IJK} t^I t^J t^K= {1\over 3!}\epsilon_{LMN} \epsilon_{PQR} t^{LP}
t^{MQ} t^{NR}=\det t.
\ee
Here the symmetric matrix  $t^{LP}= t^{PL}$ is real. The cubic invariants in
terms of real charges are
\bea
I_3(p) = {1\over 3!}\epsilon_{LMN}\, \epsilon_{PQR}\, p^{LP}p^{MQ} p^{NR}\ ,
\qquad I_3(q) = {1\over 3!}\epsilon^{LMN} \epsilon^{PQR} q_{LP}q_{MQ} q_{NR}.
\eea

\subsection{ Magic prepotentials from $d=5$ and the attractor equations}
\label{attractsec}

To connect the $d=5$ construction with real coordinates $Y^I$ to
four-dimensional holomorphic coordinates $X^I$  we have to give a prescription
for complexifying the real coordinates. We can then present the holomorphic
prepotentials $F= {I_3(X_1, X_2)\over X^0}$ of $d=4$ $N=2$ supergravities for
all of our models.  For the magic supergravities, the prescription is sweet and
simple: we just relax the reality condition on our defining matrices, keeping
it only when we need to define a complex conjugate quantity.  In the
quaternionic case, for example, we take our reality-constrained matrices $Y$
and split them once again into real matrices $Y = Y_1 + i Y_2$ satisfying $Y_1
= \Omega Y_1\Omega^T,\, Y_2 = - \Omega Y_2\Omega^T$. Regular complexification
now requires us to turn the real matrices $Y_1,Y_2$ into complex matrices
$X_1,X_2$ satisfying the linear constraint $X_1 = \Omega X_1\Omega^T,\, X_2 =-
\Omega X_2\Omega^T$. If we write $X = X_1 + i X_2$ we now have a new complex
matrix independent of any linear constraints, the new conjugate variable is
${\overline X} = X^*_1 + i X^*_2 = \Omega(X^*)\Omega^T$.

To summarize, in general we apply the standard complexification to our real
coordinates $Y_1$ and $Y_2$ to get a holomorphic coordinate $X = X_1 + i X_2$.
In the bi-index notation, what remains of the reality constraint on $Y$ is a
different definition for complex conjugation on our vector space:
\bea
\Longrightarrow && Y^{\Lambda\Sigma} \to X^{\Lambda\Sigma},\qquad \overline
{X}^{\Lambda\Sigma} = \Omega (X^{\Lambda\Sigma})^* \Omega^T.
\label{bar}\eea
Thus the ``bar'' operation is defined to be not
just complex conjugation but also a matrix multiplication as shown above.

The prepotentials for all magic $N=2$ supergravities the their cubic invariant
in five dimensions. These prepotentials may be used in future for the analysis
of the Legendre transform and relation to the generalized Hitchin functional,
as in \cite{Ooguri:2004zv},\cite{Pioline:2005vi}.
\bea
J^{\mathbb{O}}\qquad \Rightarrow \qquad F(X)= &&\frac1{3! X^0}\,
Tr( \Omega ( X_1+ i X_2)  \Omega ( X_1+ i X_2)   \Omega ( X_1+ i X_2) )=\nonumber \\
&&  \frac1{3! X^0}\,[Tr( \Omega \,   X_1 \,  \Omega \,   X_1 \, \Omega \,   X_1
)- 3 Tr( \Omega \,   X_1 \,  \Omega \,   X_2 \,   \Omega \, X_2 )]
\eea
\be
X_1= \tilde \Omega X_1 \tilde \Omega^T \ , \qquad Tr( X_1 \Omega) =0 \ , \qquad
X_2= -\tilde \Omega X_2 \tilde \Omega^T  \ , \qquad Tr( X_2 \Omega) =0 \ .
\ee
Here $X_1$ and $X_2$ are the complexified fields corresponding to the real and
imaginary parts of the antisymmetric traceless matrix $Y^{ab}$ with $Y^*= -
\Tilde \Omega Y \tilde \Omega$, $a,b=1,\dots, 8$.
\bea
J^{\mathbb{Q}}\qquad \Rightarrow \qquad F(X)= &&{1\over X^0}\, Pf\,  ( X_1+ i X_2)  =\nonumber \\
&&  {1\over 3!\cdot 2^3 X^0} \epsilon_{\Lambda \Sigma \Delta \Gamma \Pi
\Omega}\,  X_1^{\Lambda \Sigma}( X_1^{\Delta\Gamma} X_1^{\Pi \Omega}- 3
X_2^{\Delta\Gamma} X_2^{\Pi \Omega})
\eea
Here $X_1$ and $X_2$ are antisymmetric and $X_1= \Omega X_1 \Omega^T$, $X_2=
-\Omega X_2 \Omega^T$, with $\Lambda, \!\Sigma = 1..6$.
\bea
J^{\mathbb{C}}\qquad \Rightarrow \qquad F(X)= &&{1\over X^0}\, \det( X_1+ i X_2)  =\nonumber\\
&& {1\over 3!  X^0}\, \epsilon _{IJK} \epsilon_{\bar I \bar J \bar K}
X_1^{I\bar I} (X_1^{J\bar J} X_1^{K\bar K}- 3 X_2^{J\bar J} X_2^{K\bar K})
\eea
Here $X_1$ is symmetric and $X_2$ is antisymmetric.  We can put them together
in a single 3x3 matrix, $X$ in the $({\bf 3},{\bf \bar{3}})$ representation of
$H_4 = S(U(3)\times U(3))$.
\bea
J^{\mathbb{R}}\qquad \Rightarrow \qquad F(X)= {1\over X^0}\, \det X =
  {1\over 3! X^0}\epsilon_{LMN} \epsilon_{PQR} X^{LP}X^{MQ} X^{NR}\
\eea
Here $X^{LM}$  is in the symmetric representation of $H_4 = U(3)$, $L,M=1,
2,3$.

In each case, the $X^0$ field is the usual extra projective coordinate which is
a singlet of $H_5$.  We note the parallel with the five dimensional case. In
that case, the real cubic invariant was determined by a cubic invariant of the
complex form of $G_5$ plus a reality condition.  Here the holomorphic
pre-potential is determined by a cubic invariant of $H_4$ supplemented by a
different notion of complex conjugation.

 For completeness we add here the prepotentials for other $N=2$ cosets for
which the entropy formula is known. For the $L(0,P)$ models the holomorphic
prepotential is
\be
F(X) = {X\over 2 X^0}[ Y^r Y^s \eta_{rs}] \, \qquad \eta_{rs}= (1, -1, \dots ,
-1) \ , \qquad r,s, =1, \dots , P+2.
\label{LOP}\ee
The corresponding coset space in $D=4$ is ${SU(1,1)\over U(1)}\times
{SO(2,P+2)\over SO(2)\times SO(P+2)}$ and in $d=5$ we have
 $SO(1,1)\times {SO(1, P+1)\over SO(P+1)}$ and the entropy was found in \cite{Kallosh:1996tf}.

The case of the complex projective space  ${SU(1, n)\over SO(n)\times U(1)}$ is
described by a quadratic prepotential
\be
F(X) = {-i\over 4} [(X^0)^2- \sum_{i=1}^{i=n} (X^i)^2]\ ,
\label{quadratic}\ee
the entropy was found in \cite{Behrndt:1997fq}.

The BPS attractor equations for all cases can be nicely written by once more
embedding our charges into constrained complex matrices (here we use
quaternionic magic to illustrate):
\be
p^{\Lambda\Sigma} + 2i {\partial \sqrt{ J_4(p,q)} \over \partial
q_{\Lambda\Sigma}} = 2i \bar Z L^{\Lambda\Sigma}
\ee
In terms of strictly real charges $p^1,p^2$ and $q_1, q_2$ these equations can
be rewritten as:
\bea
&&p_1 +   i {\partial \sqrt{ I_4(p,q)} \over \partial q_1}= i \bar Z [L +
\Omega L \Omega^T]= 2i \bar Z L_1\\
\nonumber\\
&&p_2 +   i {\partial \sqrt{ I_4(p,q)} \over \partial q_2}= i \bar Z [L -
 \Omega L \Omega^T]= 2i \bar Z i L_2
\eea

Solving these attractor equations yields (in terms of either the constrained
complex form or split real form of the charges):
\bea
 t^{\Lambda\Sigma} = {X^{\Lambda\Sigma} \over X^0} &\rightarrow &
{p^{\Lambda\Sigma} +  2i {\partial \sqrt{ J_4(p,q)} \over \partial
q_{\Lambda\Sigma}} \over p^0+  i {\partial \sqrt{ J_4(p,q)} \over \partial
q_0}}, \qquad
 {\overline t}^{\Lambda\Sigma} = {\overline {X}^{\Lambda\Sigma}
\over \overline {X}^0} \rightarrow  {p^{\Lambda\Sigma} - 2i {\partial \sqrt{
J_4(p,q)} \over \partial q_{\Lambda\Sigma}} \over p^0 -  i {\partial \sqrt{
J_4(p,q)} \over \partial q_0}} \\
 t^{\Lambda\Sigma}_1 = {X^{\Lambda\Sigma}_1 \over X^0} &\rightarrow &
{p^{\Lambda\Sigma}_1 +  i {\partial \sqrt{ J_4(p,q)} \over \partial
q_{1\,\Lambda\Sigma}} \over p^0+  i {\partial \sqrt{ J_4(p,q)} \over
\partial q_0}},
\qquad
 {t}^{\Lambda\Sigma}_2 = { {X}^{\Lambda\Sigma}_2 \over {X}^0} \rightarrow
{p^{\Lambda\Sigma}_2 +i {\partial \sqrt{ J_4(p,q)} \over \partial
q_{2\,\Lambda\Sigma}} \over p^0 + i {\partial \sqrt{ J_4(p,q)} \over \partial
q_0}}
\label{double}
\eea
where we also exhibit the matching anti-holomorphic equation to re-emphasize
the different complex conjugation.  These equations are a slight generalization
of the one in eq. (\ref{stab2}) which allow us accommodate a more algebraic
notion of ``reality" acting on matrices and will allow us to present the black
hole composite solutions for $N=8$ and magic $N=2$ supergravities.

Finally, we can give an explicit expression for the K\"{a}hler potential of all
the magic supergravities in terms of a modified notion of ``reality" as
follows. In standard holomorphic coordinates, the K\"{a}hler potential depends
on the imaginary section.   We then get
the usual looking K\"{a}hler potential:
\be
e^{-K} = \frac i{3!}\,|X^0|^2 \, Tr\big[\Omega (t - {\bar t})\Omega (t - {\bar
t})\Omega (t - {\bar t})\big].
\ee
One has to keep in mind here that we use a definition of complex conjugation
where our anti-holomorphic special coordinates are related to the holomorphic
as follows
\be
\bar t = \Omega^T t^* \Omega
\ee
This expression for the K\"{a}hler potential again shows that it is not
invariant under K\"{a}hler transformations. One can use the gauge $X^0=1$, but
in this gauge one doesn't necessarily have $2U=K$. Moreover, it is known from
the example of the STU attractor that in this gauge $e^{-2U}$ is not equal to
$e^{K}$, see \cite{Behrndt:1996hu}. On the other hand the freedom in the choice
of the gauge for $X^0$ may be used to identify $K$ with $2U$ as in
\cite{Behrndt:1997ny}.

\subsection{ $G_6$ manifestly-symmetric   entropy of $d=4$ BPS black holes}

The explicit expressions for each case can be established  using the very
special real geometry constructions of \cite{deWit:1991nm,deWit:1992wf}. The
corresponding homogeneous special real spaces in five dimensions (very special
geometry) are
\be
{G_5\over H_5}= L(P, 1) , \qquad P=\rm dim \mathbb{A}=1,2,4,8.
\ee
in the Table 2 of \cite{deWit:1992wf}. Their corresponding K\"{a}hler spaces
with special geometry are our magic cosets ${G_4\over H_4}$ in $d=4$. We uplift
our models to  $d=6$ $(2,0)$ supergravities, following
\cite{Keurentjes:2002xc}, \cite{Andrianopoli:2004xu} (see also
\cite{Gunaydin:2005zz}). We introduce in $d=6$ a number of tensor multiplets,
$N_t=\rm dim \mathbb{A}+1$, and a number of vector multiplets $N_v=2\rm dim
\mathbb{A}$ and we split the symmetric tensor $d_{IJK}$ as follows:
\be
d_{IJK}\Rightarrow (d_{zrs}, \; d_{r\alpha \beta}) \ , \qquad
d^{IJK}\Rightarrow (d^{zrs}, \; d^{r\alpha \beta})
\ee
where $r=0, \dots, \rm dim \mathbb{A}+1$, $z=1$, $\alpha, \beta = 1, \dots ,
2\,\rm dim \mathbb{A}$. This corresponds to the split of the real
five-dimensional coordinates into $(z, b^r, a^\alpha)$ so that $z$ is the KK
singlet scalar, $b^r$ is the coordinate of the coset space ${SO(1, \rm dim
\mathbb{A}+1)\over SO(\rm dim \mathbb{A}+1)}$ and $a^\alpha$ are spinors of
$SO(1, \rm dim \mathbb{A}+1)$. We have an analogous split for the electric and
magnetic charges.
\be
p^I= (p^z, p^r, p^\alpha)\ , \qquad q_I= (q_z, q_r, q_\alpha) \ ,\qquad r=0,
\dots, \rm dim \mathbb{A}+1, \quad \alpha =1, \dots , 2\, \rm dim \mathbb{A}
\ee
For the $L(0,P)$ models one should take $r, s= 0, \dots , P+1$ and $p^\alpha =
q_\alpha =0$, following  Table 2 of \cite{deWit:1992wf}. The cubic invariants
are
\bea \label{cubic3}
I_3(p) &=& {1\over 2}\Big
(p^z\, \eta_{rs}\, p^r\,p^s  + (\gamma_{r})_{ \alpha \beta}\, p^r \,p^\alpha \, p^\beta \Big ) \\
\nonumber\\
I_3(q) &=& {1\over 2}\Big ( q_z\, \eta^{rs}\, q_r\,q_s  + (\gamma^{r})^{ \alpha
\beta}\, q_r \,q_\alpha \, q_\beta\Big )
\label{cubic4}\eea
Here $\eta_{rs}$ is the Lorentzian metric of $SO(1, \rm dim \mathbb{A}+1)$ and
$(\gamma_r)_{\alpha \beta}$ are the $\gamma$-matrices  of the groups $SO(1,\rm
dim \mathbb{A}+1)$. The details magic quaternionic supergravity are given in
Appendix A. For the groups $SO(1, 2)$, $SO(1, 3)$ and $SO(1, 9)$ we can have
real spinors and therefore the cubic invariants (\ref{cubic3}), (\ref{cubic4})
above applies to them.

The relation between  the magic exceptional octonionic $N=2$ and $N=8$
supergravity  has to do with the change from the real octonions with the norm
invariant under $SO(8)$ to the split octonions with the norm invariant under
$SO(4.4)$. In $d=6$ this means that
\bea
N=2 \qquad   &&{\rm octonionic \;  magic,
\;  real \; octonions} \qquad \Rightarrow \quad {SO(1, 9)\over SO(9)}\\
N=8 \qquad   &&{\rm octonionic,   \;  split \; octonions} \qquad \Rightarrow
\quad  {SO(5, 5)\over SO(5)\times SO(5)}
\eea

To find  the cubic/quartic  invariants of $N=8$ theory in the $G_6$ manifest
basis all we have to do it to take the cubic invariants for octonionic magic
$N=2$ in eq. (\ref{cubic3}) and (\ref{cubic4}) with the associated metric
$\eta_{rs}$ for $SO(5,5)$ instead of that for $SO(1,9)$.The $\gamma$-matrices
in the second term of cubic invariants change accordingly so that the
corresponding Clifford algebra has the  $\eta_{rs}$ signature for $SO(5,5)$
instead of $SO(1.9)$.

The case of quaternionic magic supergravity with $SO(1,5)\sim SU^*(4)$ duality
group we have to treat separately. We present this case in Appendix A.

%%%%%%%%%%%%%%%%%%%%%%%%%%%%%%%%%%
\section{BPS composites of octonionic magic $N=2$ }

 {\it Manifest $E_{6(-26)}$}

The quartic invariant of $E_{7(7)}$ is a well known invariant related
to the entropy of $N=8$ black holes.  We would like to give a similar explicit
formula for the entropy of the octonionic magic $N=2$ which is a quartic invariant
of $E_{7(-25)}$.  In this section we will extract this invariant using the
procedure given in sec.~3, but in more detail.  We will then use this
expression to write down the general multi-black hole solution
following the framework in sec.~2, with all the modifications necessary to
use our the bi-index formalism which proved so useful in deriving the
cubic and quartic invariants of octonionic magic.

Recapping the procedure in sec. 3, we introduce the charges with a $d=4
\rightarrow d=5$ split corresponding to the $E_{7(-25)}\rightarrow E_{6(-26)}
\times SO(1,1)$ split; in the $N=8$ case this is an $E_{7(7)}\rightarrow
E_{6(6)} \times SO(1,1)$ split. The real $\mathbf{56}$ of $E_{7(-25)}$ which
combines electric and magnetic charges is split into 2 real charges $p^0, q_0$
and 54 electric and magnetic charges in the ${\bf 27}'$ and ${\bf 27}$
representations of $E_{6(-26)}$, written as representations of $Usp(6,2)$. This
means we have a complex antisymmetric skew-traceless $p^{ab}= (p_1^{ab}+i
p_2^{ab})$ and $q_{ab}= (q_{1\,ab} - i q_{2\, ab})$ satisfying a reality
constraint $p^*= \tilde \Omega\, p\, \tilde \Omega^T$ and analogous for $q$.
 Here
\be
 \tilde \Omega = K^{(3,1)} \,\Omega
\ee
where the matrices $K^{(3,1)}$ and $ \Omega$ are defined in eqs. (\ref{K}),
(\ref{alpha}). Alternatively, one could use the real charges, $p_1^{ab},
p_2^{ab}$ and $q_{1\, ab}, q_{2\, ab}$, as is and lose explicit duality
symmetry.

The quartic invariants of $E_{7(7)}$ and $E_{7(-25)}$ are both
\be
J_4 (p^0, q_0, p^{ab}, q_{ab})
= -(p\cdot q)^2 + 4\Big (q_0 I_3(p) - p^0 I_3(q) +\{ I_3(q), I_3(p)\}   \Big) \ .
\label{quarticO} \ee
with cubic invariants  given by
\bea\label{cubicOp}
I_3(p)&=& \frac1{3!}\,Tr( \Omega \, p\,  \Omega \, p\,   \Omega \, p )=
\frac1{3!}\, Tr( \Omega \, p_1 \, \Omega \,   p_1 \,   \Omega \,   p_1 )
- \frac12\, Tr( \Omega \,   p_1 \,  \Omega \,   p_2 \,   \Omega \,   p_2 )\\
\nonumber\\
I_3(q)&=& \frac1{3!}\,Tr( \Omega' \, q\,  \Omega' \, q\,   \Omega' \, q )=
\frac1{3!}\, Tr( \Omega' \,   q_1 \,  \Omega' \,   q_1 \,   \Omega' \,   q_1 )-
\frac12\, Tr( \Omega' \, q_1 \, \Omega' \, q_2 \,   \Omega' \,   q_2 )
\label{cubicOq}\eea
and the scalar product of charges and the Poisson bracket of cubic
invariants are defined as follows
\bea
\label{complexdefs}
p\cdot q &\equiv& p^0 q_0 + \frac12\, p^{ab}q_{ab}= p^0 q_0 + \frac12\,
p^{ab}_1 q_{1\,ab} + \frac12 p^{ab}_2q_{2\,ab} \ ,\\\nonumber\\ \{ I_3(q),
I_3(p)\} &\equiv& 2{\partial I_3(q)\over
\partial q_{ab}} \; {\partial I_3(p)\over \partial p^{ab}} = \frac12\, {\partial
I_3(q)\over \partial q_{1\,ab} } \; {\partial I_3(p)\over \partial p^{ab}_1} +
\frac12 {\partial I_3(q)\over
\partial q_{2\,ab} } \; {\partial I_3(p)\over \partial p^{ab}_2} .
\eea
All the formulas above with complex matrices are real due to the reality
condition on the charge matrices. We note that any derivatives we write in the
octonionic case need to take into account the fact that we differentiate with
respect to skew-traceless matrices. This corresponds to restricting the tangent
space using the tracelessness condition and operationally gives the following
definition for the derivative:
\be
\frac{\partial p^{ab}}{\partial p^{cd}}= \delta^a_c \,\delta^b_d - \delta^a_d\,
\delta^b_c - \frac14 \Omega_{ab}\, \Omega_{cd}.
\ee
It would be interesting to understand exactly witch change of variable would be
necessary to match with our expression for the quartic invariant, $J_4$, with
the one in \cite{Gunaydin:2004md} written explicitly in terms of four
dimensional charges.

Overall, the only difference between the endpoint expressions for $E_{7(7)}$
and $E_{7(-25)}$ is exactly in the reality constraint on the charges, i.e.
which parts of $p$ get slotted in $p_1$ or $p_2$ by our choice of
$\tilde\Omega$:
\bea
N=8: \qquad  E_{7(7)}\rightarrow E_{6(6)} \times SO(1,1) &\Rightarrow & p=
\pm\, \tilde\Omega^T\,  p^* \,  \tilde\Omega
= \,  K^{0,4} \Omega^T\,  p^* \, \Omega K^{0,4}\\
\nonumber\\
\!\!\!\!\!\!\!\!\!\!
 N=2\;{\rm magic}:\;\; \; E_{7(-25)}\rightarrow E_{6(-26)}
\times SO(1,1) &\Rightarrow & p= \, \tilde \Omega^T \, p^* \, \tilde \Omega =
\pm\,  K^{3,1} \Omega^T\,  p^* \,  \Omega K^{3,1}.
\eea

Following sec. 2, to describe the BPS composites of octonionic magic $N=2$ we
need a set of $1+n$ fundamental $\mathbf{56}$-dimensional  representations of
$E_{7(-25)}$ . Here $n$ is the number of centers of our multicenter solution
\be
{\bf h}\equiv  (h^0, h_0, h^{ab}, h_{ab}) \ , \qquad {\bf \Gamma_s}\equiv (p^0,
q_0, p^{ab}, q_{ab})_s \, \qquad a,b=1,\dots , 8,  \qquad s=1, \dots , n.
\ee
We  introduce a 56-real-dimensional harmonic function in terms of two real
harmonics and two constrained complex matrix harmonics:
\be
\H(\vec x)= (H^0, H_0, H^{ab}, H_{ab})  = {\bf h} +\sum_{s=1}^{n} {\G_s
\over |\vec x-\vec x_s|}.
\label{harmI}\ee
Here we have introduced $n$ constant $E_{7(-25)}$ fundamentals ${\bf
\Gamma}_s\equiv (p, q)_s$ with $s=1,\dots , n$ and one constant $E_{7(-25)}$
fundamental, $h\equiv (p_{\infty}, q_{\infty})$ which is the asymptotic value
of the harmonic function.  Note that once again ``reality'' for $\H$ in this
notation means that $(H^{ab}, H_{ab})$ are complex matrices, they satisfy the
same reality condition as the corresponding charge matrices.  With the
information above, we are almost ready to write down the solution.  The final
ingredient we need is a definition of the symplectic invariant for our
solution, which is:
\be
<\G_s,\G_t> = p_s\cdot q_{t} - p_t\cdot q_s.
\label{symplecticdef}
\ee
We can now write the stationary metric for the BPS multicenter solution with
$J_4(\vec x)>0$:
\be
ds_4^2= - J_4^{-1/2}(\vec x) (dt+\vec \omega d \vec x)^2+ J_4^{1/2}(\vec x) d\vec x^2
\ee
where
\be
J_4(\vec x)\equiv J_4 \circ \H(\vec x) \qquad    \nabla \times \vec  \omega=
\langle \H, \nabla \H \rangle.
\ee
As before, we define $J_4 \circ \H(\vec x)$ by replacing the set of  charges in
eqs. (\ref{quarticO}) by the harmonic functions in eq. (\ref{harmI}). The
integrability condition for the solution is
\be
\langle \H, \triangle \H \rangle=0 \ , \qquad  \sum_{t=1}^{n}{\langle
{\bf\Gamma}_s, {\bf \Gamma}_t \rangle  \over |\vec x_s-\vec x_t|}+ \langle
{\bf\Gamma}_s, {\bf h} \rangle=0
\ee
For  the asymptotically flat geometry one has to require that $J_4(h)=1$. The
values of the vector fields (written as constrained complex
matrices again!), in spherical coordinates $(r_s, \theta_s, \phi_s)$ around each
center $\vec x_s$ are:
\be
A^{ab}= {\partial\over \partial  {H_{ab}}} \Big (\ln J_4(H)\Big ) (dt+\omega) -
\sum_s \cos \theta_s \, d\phi_s \otimes \Gamma_s^{ ab}
\label{gaugedef}
\ee
The 54 scalars of magic octonionic model represent the coset space
${E_{7(-25)}\over E_6\times SO(2)}$.  We give their solution in terms of the
holomorphic coordinates with the special complex conjugation explained in
subsection ({\bf \ref{attractsec}}):
\be
{X^{ab} \over X^0}(\vec x) ={H^{ab} +  2i {\partial \sqrt{ J_4(H)} \over
\partial H_{\, ab}} \over H^0+  i {\partial \sqrt{ J_4(H)} \over \partial
H_0}}.
\label{moduliMagic}\ee
If we want to write things in terms a purely real decomposition of ${\bf H}$,
$H^{ab} = H_1^{ab} + i H_2^{ab}, H_{ab} = H_1^{ab} - i H_2^{ab} $ with $H_1 =
H_1^* = \tilde\Omega H_1 \tilde\Omega^T$ and $H_2 = H_2^* = - \tilde\Omega H_2
\tilde\Omega^T$ we can split the equation above into two parts involving only
real harmonic functions and two separate holomorphic functions $t_1 =
\tilde\Omega t_1 \tilde\Omega^T$ and $t_2 = -\tilde\Omega t_2 \tilde\Omega_2$:
\be
t^{\Lambda\Sigma}_1 = {X^{\Lambda\Sigma}_1 \over X^0} \rightarrow
{H^{\Lambda\Sigma}_1 +  i {\partial \sqrt{ J_4(H)} \over \partial
H_{1\,\Lambda\Sigma}} \over H^0+  i {\partial \sqrt{ J_4(H)} \over
\partial H_0}}, \qquad {t}^{\Lambda\Sigma}_2 = {
{X}^{\Lambda\Sigma}_2 \over {X}^0} \rightarrow  {H^{\Lambda\Sigma}_2 + i
{\partial \sqrt{ J_4(H)} \over \partial H_{\Lambda\Sigma}} \over H^0 + i
{\partial \sqrt{ J_4(H)} \over \partial H_0}}
\ee
This completes the multicenter composite solution of the magic octonionic
supergravity.

{\it Manifest $SO(1,9)$}

By uplifting the magic octonionic supergravity to $d=6$ and breaking $
E_{6(-26)} \rightarrow SO(1,9)\times SO(1,1)$ we can present the
entropy formula using only real unconstrained charges. The set of 27
real $d=5$ coordinates includes a singlet of $SO(1,9)$ $z$, a
10-component vector $b^r$, and a chiral Majorana-Weyl 16-component
spinor $\psi^\alpha$. The quartic invariant of $ E_{7(-25)}$ in this
basis can be presented as a function of real 56 charges $(p^0, p^I;
q_0, q_I)$ where
\be p^I= (p^z, p^r, p^\alpha) \qquad q_I= (q_z, q_r,
q_\alpha)
\ee
Here $p^\alpha$ are $\mathbf{16}_L$ Majorana-Weyl
spinors of $SO(1,9)$ and $q_\alpha$ are $\mathbf{16}_R$ Majorana-Weyl
spinors of $SO(1,9)$.
\be J_4 (p^0, p^I, q_0, q_I) = -(p\cdot q)^2 +
4\Big (q_0 I_3(p) - p^0 I_3(q) +\{ I_3(q), I_3(p)\} \Big) \ .
\label{quartic6}
\ee
where the cubic invariants of the $E_{6(-26)}$ in this basis are given
by
\bea
I_3(p^I) = {1\over 2}\Big (p^z\, \eta_{rs}\, p^r\,p^s +
(\gamma_{r})_{ \alpha \beta}\, p^r \,p^\alpha \, p^\beta \Big )\qquad
I_3(q_I) = {1\over 2}\Big ( q_z\, \eta^{rs}\, q_r\,q_s +
(\gamma^{r})^{ \alpha \beta}\, q_r \,q_\alpha \, q_\beta\Big )
\label{cubic6}
\eea
and the scalar product of charges and the Poisson bracket of cubic
invariants are defined as follows
\be
p\cdot q \equiv p^0 q_0 + p^z
q_z+ p^{r} q_{r} + p^\alpha q_\alpha \ , \qquad \{ I_3(q),
I_3(p)\}\equiv {\partial I_3(q)\over \partial q_{I} }\; {\partial
I_3(p)\over \partial p^{I}}
\label{6}
\ee
Here $\eta_{rs}$ is the Lorentzian metric for the space $SO(1, 9)$ and
$\gamma$ are the chiral $\gamma$-matrices.

The relation between the entropy of the magic exceptional octonionic $N=2$ and
$N=8$ supergravity in this basis is simple. The metric $\eta_{rs}$ for $N=8$ is
that of $SO(5,5)$ instead of $SO(1,9)$ and the $\gamma$-matrices in the second
term of cubic invariants are the one for $SO(5,5)$ instead of $SO(1,9)$.

Since the entropy of exceptional magic $N=2$ supergravity in
eqs. (\ref{quartic6})-(\ref{6}) in $d=6$ basis is given in terms of
the real 56 charges we may immediately use it for the standard from of
the composite multicenter black hole solutions using equations from
the section 2.

The choice of the most appropriate basis for the solutions, the one
with manifest $E_{6(-26)}$ or with manifest $SO(1.9)$ may depend on
the problem.

%%%%%%%%%%%%%%%%%%%%%%%%%%%%%%%%%%%%%%%%%%%%%%%%%%%%%%%%%%%%%%%%%

\section{ $N=8$ BPS composites }

\subsection{Solutions via truncation to  quaternionic magic $N=2$}

 $N=8$ supergravity with its ${E_{7(7)}\over SU(8)}$ coset space
can be consistently truncated to $N=2$  quaternionic magic supergravity with a
${SO^*(12)\over U(6)}$ coset space \cite{Gunaydin:1983rk}. Hence 1/2 BPS
multi-center solutions of quaternionic magic supergravity are simultaneously 1/8 BPS
multi-center solutions of $N=8$ $d=4$ supergravity.  This fact has the useful
consequence that to generate multi-center solutions of $N=8$ we can use magic $N=2$ and
the procedure summarized in the section 2 to extend the black hole
attractor equations to stabilization equations in terms of harmonic
functions.

Our first step is to identify the proper set of 32 charges  which transforms as
a real spinor representation of the $SO^*(12)$ duality group and are truncated
from the 56-dimensional fundamental representation of the $E_{7(7)}$ duality
group of the $N=8$ model. Note that $E_{7(7)}$ decomposes into $SU(2)\times
SO^*(12)$ and therefore we can use the following split
\be
E_{7(7)} \rightarrow SU(2)\times   SO^*(12),
\qquad  \mathbf{56}\rightarrow (\mathbf{2,12}) + (\mathbf{1,32})
\ee
Thus we keep all 32 charges of $SO^*(12)$ and take as vanishing the remaining
24 charges of $E_{7(7)}$.  For this quaternionic magic model we can use the formulas
in sec. 3 to write our expressions with the $G_5= SU^*(6)$ duality manifest or with
the $G_6= SO(1,5)\sim SU^*(4)$ manifest. In both cases we naturally work with
complex coordinates in $d=5$ satisfying a reality constraint.  We will
quickly recap here the solution with manifest $G_5=SO^*(12)$ duality,
and then discuss some interesting features of the solution obtained
from the truncated $N=8$ structure.

One 32-dimensional set of charges of  $SO^*(12)$ is given by
\be
\Gamma= (p^0, p^{\Lambda \Sigma}; q_0, q_{\Lambda \Sigma})\ , \qquad \Lambda,
\Sigma =1, \dots \, 6.
\label{Gamma}
\ee
where the 15 $p^{\Lambda \Sigma}$'s satisfy the constraint $p= \Omega p
\Omega^T$ and similarly for the electric charges $q_{\Lambda \Sigma}$. These
charges transform as an anti-symmetric plus singlet of the compact subgroup
$H_5 = Usp(6)$.  As in the octonionic case, we can split our charges as
$p^{\Lambda \Sigma}= p^{\Lambda \Sigma}_1 + i p^{\Lambda \Sigma}_2$ and
$q_{\Lambda \Sigma}= q_{1\,\Lambda \Sigma} - iq_{\,2\Lambda \Sigma}$.  The
cubic invariants are
\be
I_3(p) = +{\rm Pf}\, p,\qquad I_3(q) = -\,{\rm Pf}\, q,
\ee
and the scalar product of charges and the Poisson bracket of cubic
invariants are defined as follows
\bea
\label{complexdefs2}
p\cdot q &\equiv& p^0 q_0 + \frac12\,p^{\Lambda\Sigma}q_{\Lambda\Sigma}= p^0
q_0 + \frac12 p^{\Lambda\Sigma}_1q_{1\,\Lambda\Sigma} + \frac12
p^{\Lambda\Sigma}_2q_{2\,\Lambda\Sigma} \ ,\\\nonumber\\
\{ I_3(q), I_3(p)\} &\equiv& 2{\partial I_3(q)\over \partial q_{\Lambda\Sigma}}
\; {\partial I_3(p)\over
\partial p^{\Lambda\Sigma}} = \frac12 \,{\partial I_3(q)\over \partial q_{1\,\Lambda\Sigma}}
\; {\partial I_3(p)\over
\partial p^{\Lambda\Sigma}_1} + \frac12\,{\partial I_3(q)\over
\partial q_{2\,\Lambda\Sigma}} \; {\partial I_3(p)\over \partial p^{\Lambda\Sigma}_2} .
\eea
allowing us to define $J_4(p^0,q_0,p^{\Lambda\Sigma},q_{\Lambda\Sigma})$ (if we
map $p^0$ and $q_0$ to $Z^{78}$ and $Z_{78}$ in \cite{Gunaydin:2004md} we get a
matching expression for $J_4$).

The quaternionic multi-center solution starts with $(1+n)$ of these spinor
representations of ${SO^*(12)}$. Here $n$ is the number of centers
of our multicenter solution
\be
{\bf h}\equiv  (h^0, h^{\Lambda \Sigma}; h_0, h_{\Lambda \Sigma}) \ , \qquad
{\bf\Gamma}_s\equiv (p^0, p^{\Lambda \Sigma}; q_0, q_{\Lambda \Sigma})_s \qquad
s=1, \dots , n.
\ee
The first set ${\bf h}$ defines the value of the harmonic spinor at infinity. The
other $n$ spinors are the quantized electric and magnetic charges at each center.
We can now introduce a constrained complex harmonic function:
\be
\H(\vec x)= (H^0, H^{\Lambda \Sigma}; H_0, H_{\Lambda \Sigma})  = {\bf h}+
\sum_{s=1}^{n}{{\bf \Gamma}_s \over |\vec x-\vec x_s|},
\label{harm}\ee
which gives the stationary metric for the BPS multicenter solution as
\be
ds_4^2= - J_4^{-1/2}(\vec x) (dt+\vec \omega d \vec x)^2+ J_4^{1/2}(\vec x)
d\vec x^2.
\ee
The standard definitions still apply:
\be
J_4(\vec x)\equiv J_4 \circ \H(\vec x),\qquad J_4({\bf h})= 1,
\qquad    \nabla \times \vec  \omega= \langle \H, \nabla \H \rangle,
\ee
as well as the usual integrability condition:
\be
\sum_{t=1}^{n}{\langle {\bf\Gamma}_s, {\bf\Gamma}_t \rangle  \over |\vec
x_s-\vec x_t|}+ \langle  {\bf\Gamma}_s, {\bf h} \rangle=0.
\ee
The vector fields are combined into a constrained complex matrix defined just
as in eq.(\ref{gaugedef}).  The 30 scalars of magic quaternionic model
represent the coset space ${SO^*(12)\over U(6)}$. We will give the solution for
them using our  modified attractor equations (\ref{double})
\bea
{X^{\Lambda \Sigma} \over X^0}(\vec x) &=& {H^{\Lambda \Sigma} +  2i {\partial
\sqrt{ J_4(H)} \over \partial H_{\Lambda \Sigma}} \over p^0+  i {\partial
\sqrt{ J_4(H)} \over \partial H_0}} \, ,
\eea
also written as
\bea
 {X^{\Lambda \Sigma}_1 + i X^{\Lambda \Sigma}_2 \over
X^0}(\vec x) &=& {H^{\Lambda \Sigma}_1 +  i{\partial \sqrt{ J_4(H)} \over
\partial H_{1\,\Lambda \Sigma}} \over p^0+  i {\partial \sqrt{ J_4(H)} \over
\partial H_0}} + i \, {H^{\Lambda \Sigma}_2 +i {\partial \sqrt{
J_4(H)} \over
\partial H_{2\,\Lambda \Sigma}} \over p^0+  i {\partial \sqrt{ J_4(H)} \over
\partial H_0}}
 \ .
\label{QMagic}\eea
At infinity the scalars take values defined by the harmonic function at
infinity, when $\H={\bf h}$. Near each center at $\vec x=\vec x_s$ the
attractor values of the scalars are defined as follows
\be
{X^{\Lambda \Sigma} \over X^0}(\vec x_s) ={H^{\Lambda \Sigma} +  2i {\partial
\sqrt{ J_4(H)} \over \partial H_{1\, \Lambda \Sigma}} \over p^0+  i {\partial
\sqrt{ J_4(H)} \over \partial H_0}}|_{H \Rightarrow {\Gamma_s\over |\vec x-\vec
x_s|}}.
\label{horizon}\ee
Note that if we would keep only canonical 3+3 of our 15 $p$ and $q$ we would
reproduce the Caley's Hyperdeterminant of the 2x2x2 matrix, which will also
give us the STU multicenter black hole solutions.

\subsection{On the uniqueness of the $N=8$ multi-center BPS solution}

We want to argue that the most general 1/8 BPS multicenter solution is
given by the 1/2 BPS solution of the quaternionic magic N=2
supergravity, modulo and overall $SU(8)$ rotation. Indeed, using an
$SU(8)$ rotation one can diagonalize the $N=8$ central charge matrix
so that it has only the non-vanishing complex eigenvalues $z_1=Z_{12},
z_2=Z_{34}, z_3= Z_{56}, z_4=Z_{78}$. As argued in
\cite{Ferrara:2006em} this leads to a condition at each attractor
point that \bea && z_1 z_2 + z^{*3}z^{*4}=0\nonumber\\ && z_1 z_3 +
z^{*2}z^{*4}=0\nonumber\\ && z_2 z_3 + z^{*1}z^{*4}=0
\label{attractors}\eea
The 1/8 BPS condition for the one center solution requires that
\be
  z_1= \rho_{BPS} e^{i\varphi_1} \neq 0 \qquad z_2=z_3=z_4=0 \qquad
  J_4^{BPS}=\rho_{BPS}^4>0 \qquad M_{BPS}=|z_1|
 \label{1}
\ee
    The largest central charge $z_1$ belongs naturally to a unique
$N=2$ subalgebra.  This defines a decomposition of the $N=8$
supergravity.  In the compact basis $SU(2)\times SU(6)$ under $SU(8)$
of $E_{7(7)}$
\be
\mathbf{28} \; \rm of \; SU(8) \quad \rightarrow \quad \mathbf{28}=
(\mathbf{1}, \mathbf{15}) + (\mathbf{2}, \mathbf{6}) + (\mathbf{1},
\mathbf{1})\; \rm of \; SU(2)\times SU(6)
\ee
The $1+15$ are antisymmetric and make a total of $32$ ($16$ electric and $16$
magnetic).  This is precisely like $28+28$ combine in $56$ of $E_{7(7)}$.
However, now the $32$ combine into a single chiral spinor of $SO^*(12)$, which
is the U duality group of the truncated $N=2$ theory!

In general finding $1/8$ BPS solutions of $N=8$ supergravity, $1/6$ BPS
solutions of $N=6$ supergravity, $1/4$ BPS solutions to $N=4$ supergravity
always involves a consistent truncation to the appropriate $N=2$
supergravity.  For the case of $N=8$, this gives the quaternionic
magic $N=2$ supergravity.

{\it Multi-center case}

It is plausible that for the solution to be BPS each center must be
$1/2$ BPS with respect to the same $N=2$ algebra. In such a case, we may
truncate our theory to a single copy of the magic supergravity with
vector multiplets only and ignore all the hypermultiplets, as we did in
the previous section, and we may claim that there are no other
solutions. The question which one still may like to ask here is the
following. Is it necessary that near each center the BPS solution can
only belong to the same $N=2$ supergravity?  Could it be that more
general solution has different $N=2$ parts of $N=8$ as unbroken
supersymmetry? For example, at some center instead of eq. (\ref{1}) we
may have \be z_1=z_3=z_4=0 \qquad z_2=\rho_{BPS} e^{i\varphi_1} \neq 0
\qquad J_4^{BPS}=\rho_{BPS}^4>0 \qquad M_{BPS}=|z_2|
 \label{2} \ee
If this is possible, one would not be able to provide the most general 1/8 BPS
solution of $N=8$ by truncating it to quaternionic magic $N=2$.  However, we
find this scenario extremely unlikely.

One way to resolve this issue and get a definite statement on the uniqueness/
non-uniqueness of our solution is by using the study of the Killing spinors of
1/8 BPS multicenter solutions in $N=8$ supergravity. These Killing spinors have
not been constructed so far.  We find it plausible to conjecture that such
Killing spinors have the structure seen in many other examples, namely: the
Killing spinor depends on $\vec x$ via a common factor $e^{U(\vec x)/2}$, where
the $g_{tt}=e^{2U(\vec x)}$, and a phase. The Killing spinor also satisfies
some constraint imposed globally. Thus the projector usually acts on a part of
the spinor which is $\vec x$-independent.  The reason is that the supersymmetry
transformations \cite{Cremmer:1979up}   are given by,
\begin{eqnarray}\label{grav}
\delta \Psi_{\mu A} &=&  D_\mu \epsilon _A+  Z_{AB\, \mu\nu} \gamma^\nu
\epsilon^B \ ,\\
\delta \chi _{ ABC } &=& \gamma^\mu P_{\mu ABCD}\, \epsilon^D+   Z_{[AB  \, \mu\nu} \sigma ^{\mu \nu}  \epsilon_{C]}
 \ .
\label{susy}\end{eqnarray}
They consist of two parts, the first is the gravitino in eq. (\ref{grav})
while the second is for the spin 1/2 fields in eq. (\ref{susy}).  We
deal only with the global part of the spinor for the spin 1/2 equation
while for the gravitino we have a spin connection which tells us how the
Killing spinor acquires an $\vec x$-dependent factor, which
depends on some harmonic functions. The projector, however, is the same for
all centers if the Killing spinor in $N=8$ has the same features as
the one in truncated models. If our expectation about the structure of
Killing spinors are justified, we will be able to confirm
definitely the uniqueness of our solution for 1/8 BPS multicenter
black holes in $N=8$ supergravity.

\subsection{ Alternate $N=8$ structure and the black hole mass formula}

In the material above, we have written the gauge fields, prepotential and
complex scalars using a symplectic section nicely adapted to an invariance
under the five-dimensional U-duality group $G_5$.  We will not give any
description of the gauge fields or the coset which is explicitly invariant
under $G_4$, but is useful to point out such a description is easily available
for understanding the metric.  This description uses some of the inherent
structures of the $N=8$ U-duality group, and by truncation applies to the
quaternionic magic $N=2$.  It is likely that the modification of a reality
structure yields a similar structure for the octonionic magic $N=2$, but we
will not explore this here.

The $N=8$ U-duality group in four dimensions, $G_4 = E_{7(7)}$ has a
covariant cubic map
  \cite{GN}
 from the
the fundamental representation back into itself and an invariant symplectic form.
Each $\mathbf{56}$  $(X^{ij}, X_{ij})$,
is spanned by the two antisymmetric real tensors $X^{ij}$ and $X_{ij}$ and
the action of $E_{7(7)}$ on them is realized in a standard way
 \cite{Cremmer:1979up}, \cite{GN}.

The quadratic invariant of $E_{7(7)}$, an invariant symplectic form,  can be constructed from
any two distinct  fundamentals as follows.
\be
\langle X, Y \rangle = - \langle Y,X \rangle := X^{ij} Y_{ij}- X_{ij}Y^{ij}
\ee
Note that such quadratic invariant vanishes for a single fundamental.
A triple product of any three $\mathbf{56}$  $\Big((X,Y,Z)^{ij}, (X,Y,Z)_{ij}\Big )$, gives a
trilinear map $\mathbf{56}\times \mathbf{56}\times \mathbf{56} \rightarrow \mathbf{56}$.

A unique quartic invariant can be constructed from a single fundamental in the following way. One can
build out of $X$ a triple which is different from the original $\mathbf{56}$. Afterwards one can use the
quadratic invariant for 2 distinct fundamentals: a fundamental and its triple: this  gives the standard form
of a quartic invariant via the following symplectic invariant
\be
J_4 (X)= {1\over 48} \langle  (X,X,X), X \rangle
\ee
The triple product obeys some relations
\bea
(X, Y, Z) &=& (Y, X, Z) + 2 \langle  X,Y \rangle Z\nonumber\\
(X, Y, Z) &=& (Z, Y, X) - 2 \langle  X,Z \rangle Y
\eea
A symplectic product of a fundamental and a triple satisfies the following relation
\be
\langle  (X,Y,Z), W \rangle = \langle  (X,W,Z), Y \rangle - 2 \langle  X, Z \rangle \langle  Y, W \rangle
\ee
The operations above have their origin in the Jordan algebra
structure, $J_3^{{\mathbb O}_s}$ behind the magic square
construction of $E_7$, and so carry over nicely to the duality group
based on the restriction of this algebra to $J_3^{\mathbb Q}$, the
quaternionic U-duality group $G_4=SO^*(12)$.  This allows us
to write for all magic supergravities:
\be
J_4(\vec x) = {1\over 48} <\big(\H(\vec x),\H(\vec x),\H(\vec x)\big),\H(\vec x)>.
\ee
Here the corresponding harmonic function $\H(\vec x)$ is in the fundamental
representation of $E_{7(-25)}$ for octonionic magic and in the appropriate
representation for all its truncations. In particular, in the quaternionic case
this expression provides us with the mass of the N=8 supergravity solution.  We
can efficiently extract this mass for our solution using the form for $J_4$
above, $J_4$'s fall-off at infinity, the properties of the cubic map and the
fact that $<{\bf h},\G>=0$. The ADM mass is defined via the asymptotic behavior
of the time-time component of the metric, as
\be g_{tt}= J_4^{-1/2}=1-{2M_{ADM}\over |\vec x|}+...
\ee
and we take into account that in our case
\be
<({\bf h},{\bf h},{\bf h}),\G>= <({\G},{\bf h},{\bf h}),{\bf h}>= <({\bf h},{\G},{\bf h}),{\bf h}>
=<({\bf h},{\bf h},{\G}),{\bf h}>.
\ee
This gives us a magic black hole mass formula:
\be
M_{\textrm ADM} = {1\over 48}\;<({\bf h},{\bf h},{\bf h}),\G>.
\ee

%%%%%%%%%%%%%%%%%%%%%%%%%%%%%%%%%%%%%%%%%%%%%%%%

\section{ BPS and non-BPS, $N=2$ versus $N=5,6, 8$ }
\subsection{$N=2$ attractors,  general case}
In \cite{Ferrara:2006em} simple algebraic attractor equations were derived for
regular black holes of $N=8$ supergravity. These equations have 2 solutions:
one BPS with 1/8 of unbroken supersymmetry and  one non-BPS solution with all supersymmetries broken. In the
case of the STU truncation the $N=8$ attractor equations correspond to the attractor equations of the
$N=2$ model.

Here we would like to analyse the generic attractor equations for N=2 theory and apply this analysis to BPS and non-BPS attractors of magic supergravities. For this purpose we start with a special geometry identity:
\bea
 p^\Sigma +i {\partial I_1\over \partial q_\Sigma} &=& 2i \bar Z L^\Sigma + 2i G^{j\bar j} {\cal D}_j Z {\cal D}_{\bar j} \bar L^\Sigma\nonumber\\
 q_\Sigma -i {\partial I_1\over \partial p^\Sigma}&=&  2i \bar Z M_\Sigma + 2i G^{j\bar j} {\cal D}_j Z {\cal D}_{\bar j} \bar M_\Sigma \label{general}\eea
The identity follows from the definition
\be
{\cal D}_j Z= (\partial_j +{1\over 2}K_{,j})Z
\label{D}\ee
where the central charge $Z= L^\Lambda q_\Lambda - M_\Lambda p^\Lambda$. One contracts eq. (\ref{D}) with ${\cal D}_{\bar j} \bar L^\Sigma G^{\bar j i}$ and proceed as in eqs. (40)-(45) in \cite{FK} where the BPS attractor equations were derived assuming $DZ=0$. A simplified form of these equations, which uses the  derivatives of the $I_1(p,q)$-invariant, was explained in Sec. 3.2 in \cite{Kallosh:2006bx}.
Equations (\ref{general}) supplemented by a vanishing of the first derivative of the potential, $2 \bar Z {\cal D} _i Z +i c_{ijk} G^{i\bar j} G^{k\bar k} {\cal D} _{\bar j}\bar Z {\cal D} _{\bar k}\bar Z =0$ \cite{FGK},
have 3 types of solutions.
\begin{enumerate}
  \item 1/2 BPS, well known
  \bea
 p^\Sigma +i {\partial I_1\over \partial q_\Sigma} = 2i \bar Z L^\Sigma \ , \qquad
 q_\Sigma -i {\partial I_1\over \partial p^\Sigma}=  2i \bar Z M_\Sigma \label{BPS} \ , \eea
\item Non-BPS,  $Z=0$, less known
  \bea
 p^\Sigma +i {\partial I_1\over \partial q_\Sigma} =  2i G^{j\bar j} {\cal D}_j Z {\cal D}_{\bar j} \bar L^\Sigma \, \qquad
 q_\Sigma -i {\partial I_1\over \partial p^\Sigma}=  2i G^{j\bar j} {\cal D}_j Z {\cal D}_{\bar j} \bar M_\Sigma \label{Z=0}\eea
Iff \,
$
2i G^{j\bar j} {\cal D}_j Z {\cal D}_{\bar j} \bar L^\Sigma= 2i \sum_{I=2}^4\bar Z_I {\cal D}_{\bar I} \bar L^\Sigma
$ \,
we find a solution of the following type
\be
Z_2\neq 0 \qquad Z_3=0 \qquad Z_4=0
\ee
and we can solve the attractor equations as follows
\be {p^\Lambda+i {\partial I_1\over \partial q_\Lambda}\over p^0+i
{\partial I_1\over \partial q_0}} = {\overline {\cal D}_{\hat 2} \bar
L^\Lambda \over \overline {\cal D}_{\hat 2} \bar L^0}\ , \qquad
{q_\Lambda-i {\partial I_1\over \partial p^\Lambda}\over q_0-i
{\partial I_1\over \partial p^0}}={\overline {\cal D}_{\hat 2} \bar
M_\Lambda \over \overline {\cal D}_{\hat 2} \bar M_0}
\label{stab}
\ee
\item Non-BPS, $Z\neq 0$. This case is particularly tractable for the
models with the coset spaces ${G_4\over H_4}$ of rank 3. All magic
models have rank 3. The matrix of central charge derivatives $DZ$ has
an $H_4$ symmetry:
\begin{itemize}
  \item $J^{\mathbb{O}} \qquad \Rightarrow \qquad H_4=E_6 \times U(1) $
  \item $J^{\mathbb{Q}}  \qquad \Rightarrow \qquad H_4=SU(6)\times U(1)$
  \item $J^{\mathbb{C}}  \qquad \Rightarrow \qquad H_4=SU(3)\times SU(3)\times U(1)$
  \item $ J^{\mathbb{R}} \qquad \Rightarrow \qquad H_4=SU(3)\times U(1)$
\end{itemize}
With an $H_4$ rotation we can skew-diagonalize the matrix $DZ$ so that it has 3
eigenvalues, $\bar {\cal D}_I \bar Z \equiv Z_I, \, I=2,3,4$. Together with
$N=2$ central charge $Z= -i Z_1$ we have 4 charges,
\be Z_1= i Z \ , Z_I= \bar D_{\hat I} \bar Z \ee The
extremality condition is given by
 \be 2 \bar Z {\cal D}_{\hat I}Z + i
d_{\hat I \hat J \hat K}\bar Z_{\hat {\bar J}} \bar Z_{\hat {\bar K}}
\delta^{\hat J \hat {\bar J}}\delta^{\hat K \hat {\bar K}}=0
\ee
and it becomes for magic models in normal frame
\bea && Z_1 Z_2 +
Z^{*3}Z^{*4}=0\nonumber\\ && Z_1 Z_3 + Z^{*2}Z^{*4}=0\nonumber\\ &&
Z_2 Z_3 + Z^{*1}Z^{*4}=0
\eea
exactly as in $N=8$ case \cite{Ferrara:2006em}. The solution is as in \cite{Ferrara:2006em}
\be
Z_1= \rho e^{i(\pi-3\phi)} \ , \qquad Z_I= \rho e^{i\phi}, \qquad Z_1
Z_2 Z_3 Z_4= \rho^4 e^{i\pi}=-\rho^4
\ee
Here we have satisfied the attractor equations: $Z_I Z_J + Z^*_K Z^*_M =0,\qquad I\neq J\neq
K\neq M$.
We may use this in an identity (\ref{general}) and we find in case that $\phi=0$
\bea
p^\Sigma +i {\partial I_1\over \partial q_\Sigma} &=& 2\rho
( L^\Sigma + i \sum_{\hat {\bar I}} \bar {\cal D}_{\hat {\bar I}} \bar L^\Sigma)\\
\nonumber\\
q_\Sigma -i {\partial I_1\over \partial p^\Sigma}&=&   2\rho ( M_\Sigma + i
\sum_{\hat {\bar I}} \bar {\cal D}_{\hat {\bar I}} \bar M_\Sigma) \label{magic}
\eea
and
\bea
{p^\Sigma +i {\partial I_1\over \partial q_\Sigma}\over p^0 +i {\partial
I_1\over \partial q_0}} =  { L^\Sigma + i \sum_{\hat {\bar I}} \bar {\cal D}_{\hat {\bar I}} \bar
L^\Sigma\over L^0 + i \sum_{\hat {\bar I}} \bar {\cal D}_{\hat {\bar I}} \bar L^0 } \ , \qquad
{q_\Sigma -i {\partial I_1\over \partial p^\Sigma}\over q_0 -i {\partial
I_1\over \partial p^0}}=    { M_\Sigma + i \sum_{\hat {\bar I}} \bar {\cal D}_{\hat {\bar I}} \bar
M_\Sigma\over  M_0 + i  \sum_{\hat {\bar I}}\bar {\cal D}_{\hat {\bar I}} \bar M_0    }
\label{magic2}
\eea
\end{enumerate}
Equations (\ref{magic2}) define the values of moduli fields in terms of charges in the non-BPS case. We have verified \cite{KSS}  that these equations provide a non-BPS attractor solution of the STU model in example studied in \cite{Tripathy:2005qp}.

\subsection{ $N=2$ versus $N= 5, 6, 8$ and BPS versus ,non-BPS}

The extended supergravities $N= 5, 6$ as well as $N=8$ have no matter
multiplets, only the gravitational one. Therefore they are very restricted. BPS
black holes in these theories have been studied before in
\cite{Andrianopoli:1996ve}. Since now we understand certain features on $N=8$
BPS and non-BPS black holes and their relation to $N=2$ BPS and non-BPS ones,
as shown in \cite{Ferrara:2006em}, we may extend this relations also to include
$N=6$ and $N=5$ theories.

The $N=8$ supergravity has a consistent truncation both to $N=6$ and $N=2$ supergravity theories, depending on whether one keeps six or two of the eight gravitinos. In both cases one ends up with $J^Q$ based on the ${SO^*(12)\over U(6)}$ manifold. This manifold, because it is consistent with N=2, is indeed a special K\"{a}hler symmetric space. The charge vector is in the $\mathbf{32}$ of $SO^*(12)$ (chiral spinor). From an $N=2$ point of view
\be
\mathbf{32}= \mathbf{15}+ \overline {\mathbf{15}}+ \mathbf{1}+ \overline {\mathbf{1}}
\ee
where $\mathbf{15}$ are in the matter multiplet and $\mathbf{1}$ is the graviphoton.

In $N=6$ supergravity the central charge is $Z^{AB}= - Z^{BA}$ with $A,B= 1, ..., 6$ and there is also a singlet charge $Z$ \cite{Andrianopoli:1996ve}. The black hole potential is
\be
V_{BH}= {1\over 2} Z_{AB}  \bar Z^{AB} + Z \bar Z \ ,
\ee
The 1/6 BPS solution of N=6 theory was identified in
\cite{Andrianopoli:1996ve}: it has $Z=0$ and a skew-diagonal $Z_{AB}$ with
$Z_{12}\neq 0$ and $Z_{34}= Z_{56}=0$. From the $N=2$ point of view such solution
is a non-BPS solution (with vanishing central charge). On the other hand, the
1/2 BPS  solution of $N=2$ theory with $Z\neq 0$ and $Z_{AB}=0$ corresponds to
a non-BPS solution of the N=6 theory (with vanishing $N=6$ central charge).

Let us now comment on $N=5$ theory and relation of 1/5 BPS solution to 1/2 BPS of the relevant $N=2$ theory. The $N=5$ special K\"{a}hler geometry theory is based on the $SU(5,1)\over U(5)$ symmetric space with 5 complex scalars. The central charge $Z_{AB}$ has a non-vanishing component $Z_{12}\neq 0$ whereas  $Z_{34}= 0$ is vanishing: this is a 1/5 BPS solution.
Despite the fact that the $N=5$ model has the same sigma model as the corresponding $N=2$ theory, the vector part of these models is different: $N=2$ has six vectors and $N=5$ has ten vectors. Therefore, as different from $N=6$ case, the 1/5 BPS solutions of $N=5$ are different from the 1/2 of $N=2$ supergravity based on $SU(5,1)\over U(5)$ symmetric space.

\section{Discussion}

    We have given a complete description of the cubic and quartic invariants
for magic supergravities.  Using these we were able to write down multi-center
1/2 BPS solutions for the exceptional octonionic magic supergravity with a
56-component charge vector at each center.  This model has  $E_{7(-25)}$
duality symmetry, which is slightly different from the duality symmetry of
$N=8$ theory, $E_{7(7)}$. We have also demonstrated how multi-center
quaternionic magic solutions with 32-component charge at each center  can be
used to describe a broad (possibly complete) array of $N=8$ 1/8 BPS
multi-center solution.

In \cite{Ferraraetal}, it was shown that any single center 1/8 BPS solution of
$N=8$ could be rotated using an $E_{7(7)}$ duality transformation to a form
involving just a single center solution the N=2 STU truncation of $N=8$. For
the non-BPS attractors of magic supergravities we have made used of this
truncation to the STU model. Using the solutions of the non-BPS attractors for
$N=8$ supergravity presented in \cite{Ferrara:2006em}  we have found the
general solutions of the non-BPS attractors for the $N=2$ STU model.

  The duality rotation which takes a single center solution to an STU form is not
powerful enough for a generic 1/8 BPS solution with multiple centers: such a
rotation is clearly not enough to rotate every center to STU form.  For
solutions generated via quaternionic magic supergravity, it would be
interesting to work out which is the minimal $N=2$ truncation that admits
generic cases with two, three, four etc... centers, just as STU supergravity is
minimal truncation of $N=8$ which describes all single center solutions.

    Clearly, $N=2$ truncations play a major role in understanding 1/8 BPS
states of $N=8$ supergravity.  Apart from clarifying if $N=2$ truncation is the
only way to achieve 1/8 BPS configurations, a clearer look at these truncations
in context of black holes might be productive.  Different truncations allow a
varying number of hypermultiplet scalars (see e.g. \cite{Andrianopoli:2001zh})
from the $N=8$ coset to survive. The hyper scalars vevs typically play a
spectator role, since they will not vary in BPS black hole solutions, but for
an observer near infinity they may reveal some information about just which
$N=2$ truncation is in operation for a given multi-center solution.

For future reference, we summarize the hyper-multiplet scalars which the
survive from the $N=8$ coset in the truncation to the various magic
supergravities (for details see \cite{Andrianopoli:2001zh}).  In quaternionic
magic, we have $E_{7(7)} \to SO^*(12)\times SU(2)$ and no hypers survive.  In
the complex case, $E_{7(7)} \to SU(3,3) \times SU(2,1)$ and we get four hyper
scalars parameterizing the coset $SU(2,1)/SU(2)\times U(1)$; the real case
corresponds to $E_{7(7)} \to Sp(6,{\mathbb R}) \times G_{2(2)}$ and has eight
hyper scalars, $G_{2(2)}/SO(4)$.  The STU truncations has hyper scalars in
$SO(4,4)/SO(4)\times SO(4,4)$ (see \cite{Ferraraetal}).  As we restrict the
vector moduli space, the hyper moduli space increases in multiples of four (it
is quaternionic).

    Finally, given the relationship between extremal non-BPS and BPS solutions of $N=2$
magic quaternionic supergravity and those of $N=6$, it would be interesting to
understand just how many such links exists between various extremal solutions
of different supergravities, perhaps along lines related to
\cite{Kallosh:2006bx}.

{ {\bf Acknowledgments}}

It is a pleasure to thank R. D'Auria, V.~Balasubramanian, O.~Ganor, M.
G\"{u}naydin, T.~Levi, T. Ort\'{\i}n, B. Pioline, A.~Sen, A. Tomasiello and M.
Trigiante for useful conversations.  We are grateful to participants of 2006
Frascati Winter school on Attractor mechanism for the interest to this work.
The work of S.F.~has been supported in part by the European Community Human
Potential Program under contract MRTN-CT-2004-005104 ``Constituents,
fundamental forces and symmetries of the universe'', in association with INFN
Frascati National Laboratories and by D.O.E.~grant DE-FG03-91ER40662, Task C.
The work of EG was supported by the US Department of Energy under contracts
DE-AC03-76SF00098 and DE-FG03-91ER-40676 and by the National Science Foundation
under grant PHY-00-98840.  The work of R.K. was supported by NSF grant
PHY-0244728.

\section*{Appendix A: $d=6$ duality and quaternionic magic model}

This model even if we break the manifest $G_5$ duality down to $G_6$ duality still does not give us a very special real geometry in $d=5$ and the relevant complexified special geometry in $d=4$. The reason is that the spinors of $SO(1,5)\sim SU^*(4)$ as opposite to other magic models with  $SO(1,2)$, $SO(1,3)$, $SO(1,9)$ do not have real representations. Thus we have to use complex coordinates constrained by reality conditions.

Both the vectors  and the spinors are complex and satisfy the reality condition. For vectors of $SU^*(4)$ we use an antisymmetric matrix $V_{AB}=-V_{BA}$ and $V^*= -\Omega V\Omega$ where $\Omega= -\Omega^t$, $\Omega^2= -\mathbf{1}$. Thus $V_1={\rm Re} V= -\Omega {\rm Re} V \Omega$ and $V_2={\rm Im} V= \Omega {\rm Im} V \Omega$. We also introduce $\tilde V^{AB}\equiv {1\over 2} \epsilon^{ABCD} V_{CD}$. Thus the first term  in the cubic invariant, $z [(b^0)^2- \sum_{i=1}^{i=5}( b^i)^2]$ can be rewritten as $z V_{AB}\tilde V^{AB} = z (V_{1\, AB}\tilde V_1^{AB} - V_{2\, AB}\tilde V_2^{AB})$ in terms of real entries only. For spinors $\Psi_{Ai}$ of $SU^*(4)\times SU(2) $ with $A=1,\dots, 4, i=1,2$ there is a reality condition $\Psi^*_{Ai}= \Omega_{AB}\epsilon_{ij} \Psi_{Bj}$
and $\epsilon^2=-1$. We also split these spinors into real and imaginary parts which we call $\Psi_1$ and $\Psi_2$ respectively, $\Psi_1=\Omega \epsilon \Psi_1$ and  $\Psi_2=-\Omega \epsilon \Psi_2$. The second term of the cubic invariant can now be presented as a function of only real coordinates, $[V_{1\, AB}(\Psi_1^{Ai} \Psi_1^{Bj} - \Psi_2^{Ai} \Psi_2^{Bj})- V_{2\, AB} (\Psi_1^{Ai} \Psi_2^{Bj} + \Psi_1^{Bj} \Psi_2^{Ai})]\epsilon_{ij}$.

Now we need the cubic invariants in terms of charges. We take
\be
p^z, p^r, p^\alpha \Rightarrow p^z, p_{AB}, p^{Ai} \ , \qquad q_z, q_r, q_\alpha \Rightarrow q_z, q^{AB}, q_{Ai} \ , \qquad p_{AB}= - p_{BA} \, \quad q^{AB}= - q^{BA}
\ee
and \,
$
p\cdot q= p^z q_z+ p_{AB} q^{AB} + p^{Ai} q_{Ai} \, \qquad \tilde p^{AB} = {1\over 2} \epsilon^{ABCD} p_{CD} \ , \qquad \tilde q_{AB} = {1\over 2} \epsilon_{ABCD} q^{CD}
$

The cubic invariants are
\bea
I_3(p) &&={1\over 2}\Big ( p^z p_{AB} \tilde p^{AB} + p_{AB} p^{Ai} p^{Bj} \epsilon_{ij}\Big )\\
\nonumber\\
I_3(p) &&= {1\over 2}\Big ( q_z q^{AB} \tilde q_{AB} + q^{AB} q_{Ai} q_{Bj} \epsilon^{ij}\Big )
\eea
where the reality conditions are
\be
p^* = -\Omega p \Omega \ , \qquad p^{*Ai} = (\Omega \epsilon p)^{Ai}
\label{re}\ee
 As the consequence of the reality conditions (\ref{re}) the cubic invariants are real
\be
I_3^* (p) = I_3(p^*) = I_3(p) \,
\ee
and the same for electric charges. We may split the charges into real and imaginary parts
\be
p_{AB} =p_{1\, AB}  +i p_{2\, AB}\qquad p^{Ai}_1 +i p^{Ai}_2
\ee
 and find for $I_3(p)$
\bea
{1\over 2} \Big (p^z( p_{1\, AB} \tilde p_1^{AB} - p_{2\, AB} \tilde p_2^{AB})  + [p_{1\, AB}( p_1^{Ai} p_1^{Bj}- p_2^{Ai} p_2^{Bj})- p_{2\, AB}( p_1^{Ai} p_2^{Bj}+ p_2^{Ai} p_1^{Bj})]\epsilon_{ij} \Big)
\eea
and analogous for $I_3(q)$.

\end{document}